\documentclass{aa}  

\usepackage{graphicx}
\usepackage{txfonts}
\usepackage{subcaption}
\usepackage{lscape}
\usepackage{placeins}

\usepackage{natbib}
\bibpunct{(}{)}{;}{a}{}{,}

\usepackage{tabularx}
\newcolumntype{Y}{>{\centering\arraybackslash}X}
\usepackage{booktabs}
\usepackage{makecell}

\begin{document}

   \title{The compact neutron star in 4U 1746-37 revisited:\\Reassessing the mass and radius}

   \author{Kwang Hyun Sung\inst{1}\fnmsep\thanks{These authors contributed equally to this work and should be considered joint first authors.}
   	\and Myungkuk Kim\inst{2}\fnmsep\protect\footnotemark[1]
	\and Young-Min Kim\inst{3,4}
	\and Kyujin Kwak\inst{1}\fnmsep\thanks{Corresponding author; \email{kkwak@unist.ac.kr}}
	\and Chang-Hwan Lee\inst{5}}
   
   \institute {Department of Physics, Ulsan National Institute of Science and Technology, Ulsan 44919, Korea
            \and Center for Exotic Nuclear Studies, Institute for Basic Science, Daejeon, 34126, Korea
            \and Korea Astronomy and Space Science Institute, Daejeon, 34055, Korea
            \and Department of Astronomy and Space Science, University of Science and Technology, Daejeon 34113, Republic of Korea
            \and Department of Physics, Pusan National University, Busan, 46241, Korea}

   \date{Received 29 September 2025 / Accepted 16 April 2026}

   \authorrunning{K. Sung et al.}

  \abstract
   {A recent analysis of photospheric radius expansion X-ray bursts from the low-mass X-ray binary 4U 1746-37 reported unusually small mass and radius
   estimates for the neutron star, suggesting it could be a quark star or quark-cluster star.
   Here, we propose an alternative interpretation: the star's mass and radius could be underestimated from significant blocking of the X-ray flux.}
   {By introducing a blocking factor to account for the systematic reduction of observed flux relative to the intrinsic emission from the neutron star's
   photosphere, we investigated whether the reduction in observed flux can reconcile anomalous mass--radius estimates with canonical neutron star properties.} 
   {We defined the blocking factor as the fraction of the neutron star photosphere obscured from view, which scales both the observed touchdown flux and the effective emitting area.
   We solved the modified photospheric radius expansion equations analytically, which yields two distinct mathematical branches of mass–radius solutions, and employed
   Monte Carlo simulations for high-blocking scenarios.}
   {Significant blocking factors ($\mathcal{B} \gtrsim 0.8$, reducing the observed flux to $\sim17\%$ of the intrinsic emission) permit neutron star parameters consistent with
   the canonical values: $M = 1.59 \pm 0.69\,M_{\sun}$, $R = 13.0 \pm 5.45\,\mathrm{km}$, or $M = 2.12 \pm 1.08\,M_{\sun}$, $R = 9.80 \pm 4.13\,\mathrm{km}$.
   The blocking factor, which varies with the photospheric radius, provides a natural explanation for the anomalously large peak-to-touchdown flux ratio ($\sim2.0$) and
   highlights the importance of accounting for geometric system configuration in neutron star mass--radius estimates.}
   {}


   \maketitle
   \nolinenumbers

\section{Introduction}
With densities comparable to those of atomic nuclei, neutron stars are among some of the most extreme objects in astrophysical environments \citep{Piekarewicz2023}.
The concept of these ultra-dense objects was first proposed by \citet{Baade1934}, just two years after the discovery of the neutron by \citet{Chadwick1932},
when they put forward the idea that ordinary stars could collapse and transform into neutron stars through supernova events.
Neutron stars serve as cosmic laboratories for investigating the behavior of matter under extraordinarily high density and pressure,
conditions that are impossible to recreate in Earth-based experiments.
Accurately measuring their masses and radii is crucial for understanding the properties of ultra-dense matter and
constraining the equation of state (EoS) describing it, as these measurements directly probe the fundamental physics of matter
at nuclear densities and beyond \citep{Baym1971,Shapiro1983,Lattimer2001,Lattimer2007,Ozel2016}.

Type I X-ray bursts (XRBs) exhibiting photospheric radius expansion (PRE) in low-mass X-ray binary (LMXB) systems provide a powerful method for determining
neutron star masses and radii. These thermonuclear explosions occur on neutron star surfaces as hydrogen and helium accumulate from companion stars through
accretion. When the accreted material reaches sufficient temperatures and densities, runaway thermonuclear fusion produces bright X-ray flashes lasting
tens to hundreds of seconds. During particularly energetic bursts, the luminosity can reach the Eddington limit, causing the photosphere to expand due to
radiation pressure. This PRE phase exhibits characteristic signatures: increased emitting area, decreased observed temperature, and approximately constant
bolometric flux at the Eddington limit. Subsequently, as nuclear burning subsides, the photosphere contracts back to the neutron star surface, defining the touchdown phase.

In PRE-burst work, the inferred mass--radius constraints depend on the burst-analysis methodology. One approach, known as the touchdown method, utilizes the flux
measured at this specific moment (as a proxy for the Eddington limit) combined with the effective emitting area to constrain the mass and radius \citep{Lewin1993,Ozel2006,Galloway2008b,Ozel2009,Steiner2010}. Another approach, the cooling tail method, instead models the full spectral evolution of the burst decay
to determine these parameters \citep{Suleimanov2011ApJ,Nattila2017}. Collectively, these techniques have been applied to several neutron stars in LMXBs over the past decade,
providing valuable insights into neutron star properties and dense matter EoSs.

Observational advances have been driven by increasingly sophisticated X-ray missions.
The Rossi X-ray Timing Explorer (RXTE) has been instrumental through its large effective area and high temporal resolution, enabling detailed studies of
burst profiles, spectral evolution, and oscillations across an extensive sample of sources. \citet{Galloway2008} presented a comprehensive catalog of over
1,000 X-ray bursts observed by RXTE from 48 sources, providing a rich dataset for systematic studies of bursting behavior and neutron star properties.
Additional X-ray missions such as the Neutron star Interior Composition Explorer (NICER), the Nuclear Spectroscopic Telescope Array (NuSTAR), and the Neil Gehrels
Swift Observatory X-Ray Telescope (Swift-XRT) have inherited and expanded upon RXTE's legacy in X-ray astronomy through enhanced spectral, timing, and
observational capabilities. Because each instrument specializes in different areas, these modern observatories address RXTE's limitations while advancing specific aspects
of burst studies and accretion disk interactions \citep{Bradt1993,Burrows2005,Harrison2013,Gendreau2016}.

Within this observational context, we are particularly interested in 4U 1746-37 among the catalog of X-ray burst sources.
Located in the globular cluster NGC 6441, this LMXB has been extensively studied since \citet{Giacconi1974} first detected its persistent X-ray emission.
\citet{Li1977} reported the first detection of type I X-ray bursts from 4U 1746-37 using the Small Astronomy Satellite 3 (SAS-3) X-ray Observatory.
\citet{Sztajno1987} conducted an analysis of 12 hours of continuous observations from the European X-ray Observatory Satellite (EXOSAT)
and found that the XRB emissions from 4U 1746-37 are likely anisotropic, meaning that the radiation is not emitted uniformly in all directions.
\citet{Parmar1989} marked the first detection of X-ray dips from 4U 1746-37. This detection later proved crucial in classifying 4U 1746-37 as a dipping source,
a subclass of LMXBs characterized by high orbital inclinations.
\citet{Sansom1993} estimated the orbital period of 4U 1746-37 to be 5.7 hr based on the dip recurrence using data from the Ginga satellite.
\citet{Deutsch1998} reported the probable detection of the optical counterpart of 4U 1746-37 based on observations made with the Hubble Space Telescope (HST).
\citet{Jonker2000} discovered $\sim$ 1 Hz quasi-periodic oscillation (QPO) in the persistent emission and during type I X-ray bursts of 4U 1746-37.
They also found evidence that 4U 1746-37 exhibits spectral and timing properties similar to atoll sources.
\citet{Kuulkers2003} measured the distance to NGC 6441, the globular cluster hosting 4U 1746-37, to be 11.0$\genfrac{}{}{0pt}{1}{+0.9}{-0.8}$ kpc
by utilizing the luminosity level of the horizontal branch as a standard candle.
\citet{Balucinska2004} refined the measurement of the orbital period to 5.16 $\pm$~0.01 hr using long observations made with RXTE.
\citet{Galloway2008} reported 30 bursts from 4U 1746-37 with some exhibiting evidence of PRE.
\citet{Munoz2014} described 4U 1746-37 as a persistent source that predominantly remains in a thermal-dominated state (soft state),
occasionally transitioning to a hard state at low count rates.
\citet{Li2015} performed an analysis of three PRE bursts from 4U 1746-37 observed by RXTE. 
Their findings suggest the potential existence of an ultra-low-mass neutron star within this system.
\citet{intZand2019} reported three X-ray bursts from 4U 1746-37 with the Swift-XRT.
\citet{Panurach2021} analyzed quasi-simultaneous radio and X-ray observations from the Swift-XRT and the Milky Way ATCA VLA Exploration of Radio Sources in Clusters
(MAVERIC) surveys. A key finding was the absence of radio continuum emission in 4U 1746-37,
which may indicate either faint radio emission from the source or a potential quenching of its jet activity.

Particularly significant and controversial results emerged from \citet{Li2015}, whose analysis of three RXTE-observed PRE bursts yielded remarkably low mass and radius estimates.
They derived two distinct solutions depending on how they interpreted the high ratio of the peak flux to the touchdown flux ($F_{p}/F_{TD} \approx 2.0$).
In their first scenario, they attributed the high peak flux to reflection from the far-side accretion disk and assumed the observed touchdown flux represented the true Eddington limit;
this yielded the lower estimate of ($M \approx 0.21\,M_{\odot}$, $R \approx 6.3$ km). In their second scenario, they assumed the system was subject to obscuration and corrected
the touchdown flux by the factor $F_{p}/F_{TD}$, resulting in the slightly higher estimate of ($M \approx 0.41\,M_{\odot}$, $R \approx 8.7$ km). Both values fall dramatically below
canonical neutron star parameters ($\sim 1.4\,M_{\odot}$, $\sim 10$ km), leading to interpretations involving exotic compact objects such as quark stars or quark-cluster stars.
These findings challenge our understanding of dense matter physics and suggest the possible existence of ultra-low-mass neutron stars.

However, the geometric configuration in LMXB systems can substantially influence observed X-ray properties. 
Anisotropic emission patterns can arise from several key factors: nonspherical accretion disk geometry creating preferential emission directions,
obscuration as disk material blocks X-ray flux along specific sight lines, and reflection processes where photons interact with disk material before
reaching the observer. These geometric dependencies introduce systematic variations in observed flux and spectral characteristics, with angular radiation
distributions varying significantly based on viewing angle and orbital phase. Consequently, the apparent X-ray properties of a neutron star can differ
substantially depending solely on the observer's perspective relative to the binary system orientation, making geometric considerations essential for
accurate interpretation of LMXB observations and the underlying physics of these compact systems.

The classification of 4U 1746-37 as a dipping source indicates high orbital inclination where the accretion disk periodically obscures the central object.
This nearly edge-on configuration presents fundamental observational challenges: Partial neutron star obscuration causes observed thermonuclear burst properties
to vary based on the geometrical configuration rather than the intrinsic physics of the burst. While previous studies considered obscuration and reflection effects during
peak luminosity phases of PRE bursts \citep{Lapidus1985,Fujimoto1988,He2016}, they have not adequately addressed the possibility that significant flux reduction
persists even at touchdown when the photosphere contracts to its minimum size.

We propose that substantial geometric blocking of photospheric emission throughout the entire burst evolution could reconcile the unusual mass and radius
estimates for the neutron star in 4U 1746-37 with canonical neutron star values. This geometric obscuration could arise from various components of the system,
including but not limited to accretion disk and other structural elements that could obstruct the direct line of sight to the neutron star surface.
To address this potential observational bias, we introduced a blocking factor quantifying flux reduction from the observer's perspective due to line-of-sight
impediments as a simple first attempt to account for geometric effects and re-estimate the true burst flux. While traditional anisotropy factors describe
intrinsic, angular distribution of emission from the source, our blocking factor emphasizes the observational consequences of geometric obscuration,
prioritizing how much the measured flux is diminished rather than describing the overall emission pattern of the source.

For this work, we tested whether viewing-angle-dependent blocking can reconcile the unusually low mass and radius estimates of \citet{Li2015} with canonical neutron star properties.

\section{Blocking factor}
\begin{figure*}[!tb]
   \centering
   \begin{subfigure}{0.40\textwidth}
      \centering
      \includegraphics[width=\linewidth]{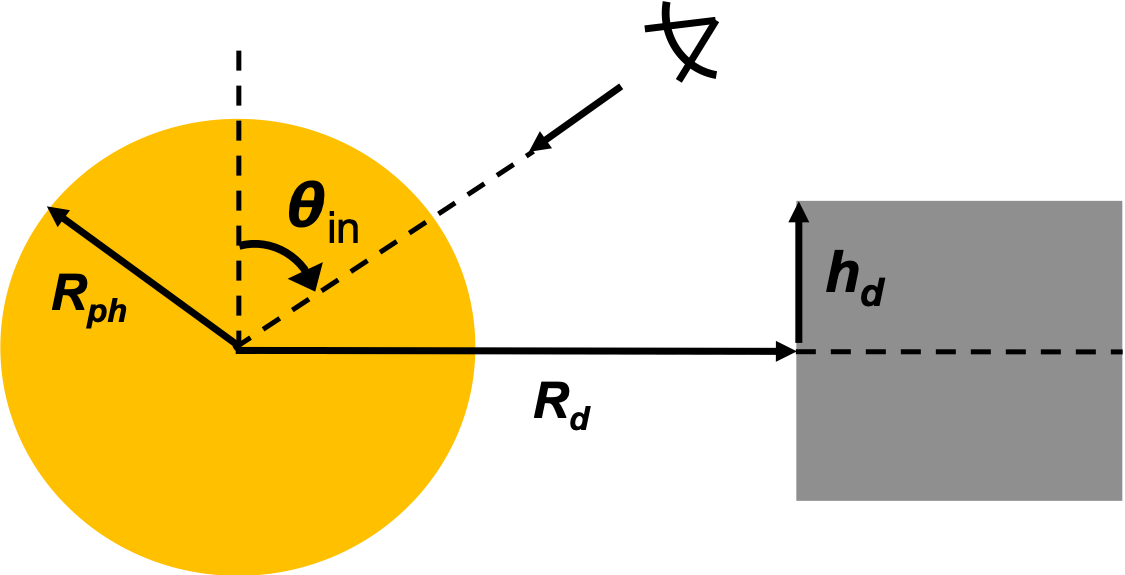}
      \label{fig_model_thin}
   \end{subfigure}
   \hspace{0.06\textwidth}
   \begin{subfigure}{0.45\textwidth}
      \centering
      \includegraphics[width=\linewidth]{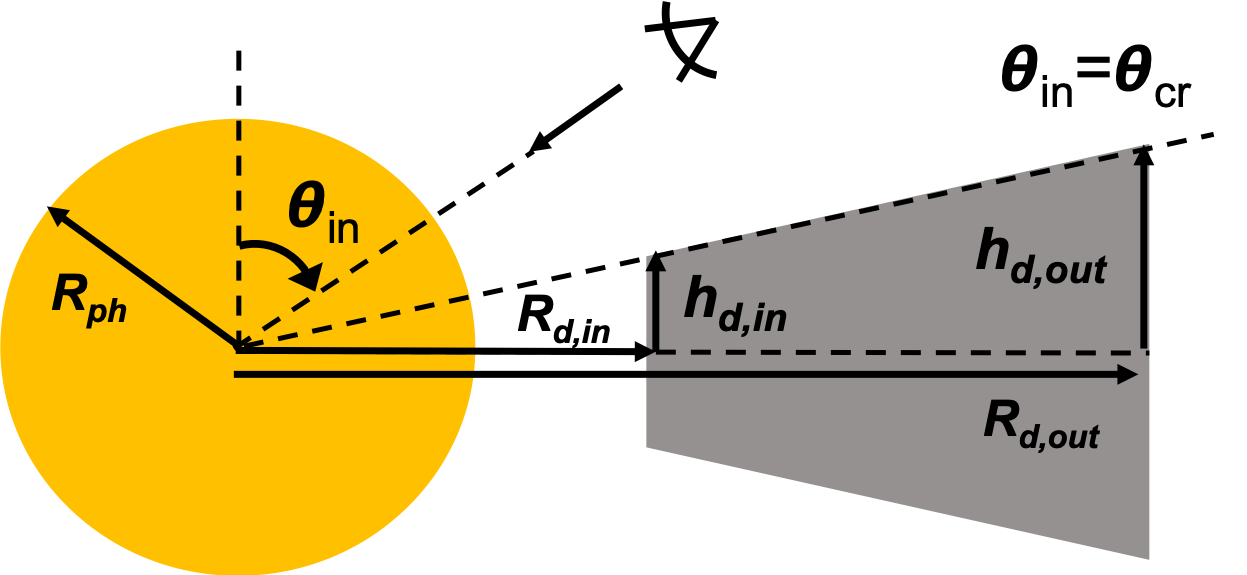}
      \label{fig_model_thick}
   \end{subfigure}
   \caption{Schematic cross-sectional view of the blocking mechanism in high-inclination LMXB systems.
         Left panel: Thin accretion disk configuration. Right panel: Thick-disk geometry. Both show a neutron star viewed by a distant observer at
         inclination angle $\theta_{\rm {in}}$, measured from the disk normal. The blocking factor $\mathcal{B}$ represents the fraction of the neutron star's
         photosphere obscured, which varies with $\theta_{\rm {in}}$ and the photospheric radius $R_{ph}$.
         Diagrams are not to scale for clarity; the actual disks extend radially much farther.}
   \label{fig1_disks}
\end{figure*}

\begin{figure*}[!tb]
   \centering
   \begin{subfigure}{0.49\textwidth}
      \centering
      \includegraphics[width=\linewidth]{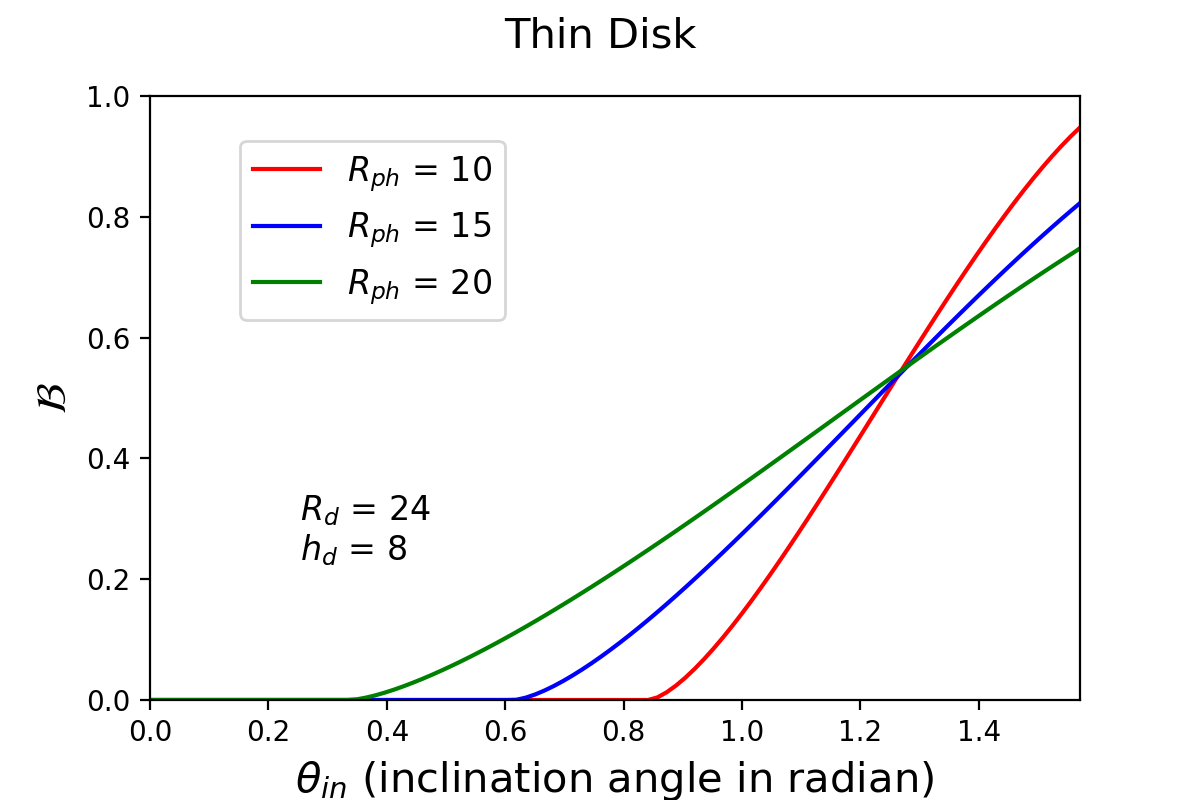}
      \label{alpha_vs_theta_thindisk}
   \end{subfigure}
   \begin{subfigure}{0.49\textwidth}
      \centering
      \includegraphics[width=\linewidth]{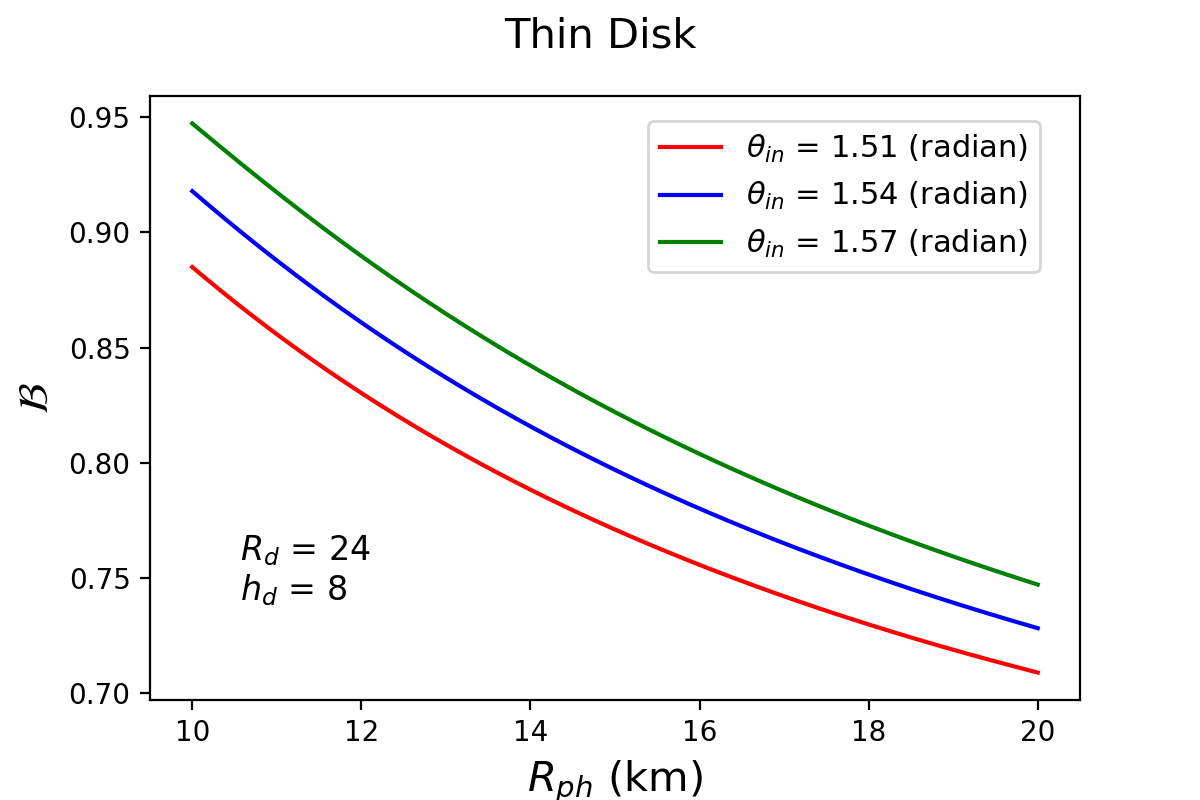}
      \label{alpha_vs_Rph_thindisk}
   \end{subfigure}
\\
   \begin{subfigure}{0.49\textwidth}
      \centering
      \includegraphics[width=\linewidth]{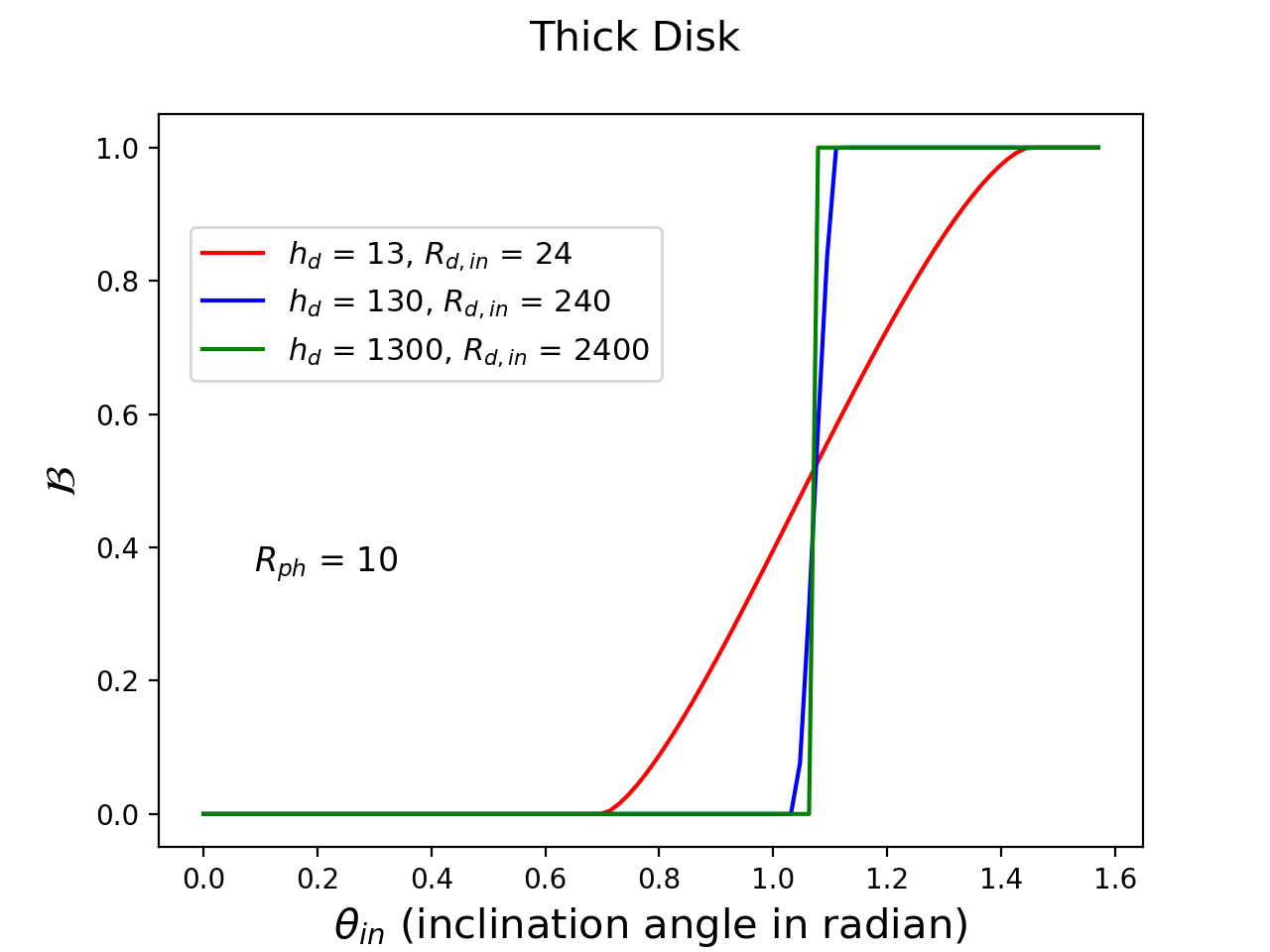}
      \label{alpha_vs_theta_thickdisk}
   \end{subfigure}
   \begin{subfigure}{0.49\textwidth}
      \centering
      \includegraphics[width=\linewidth]{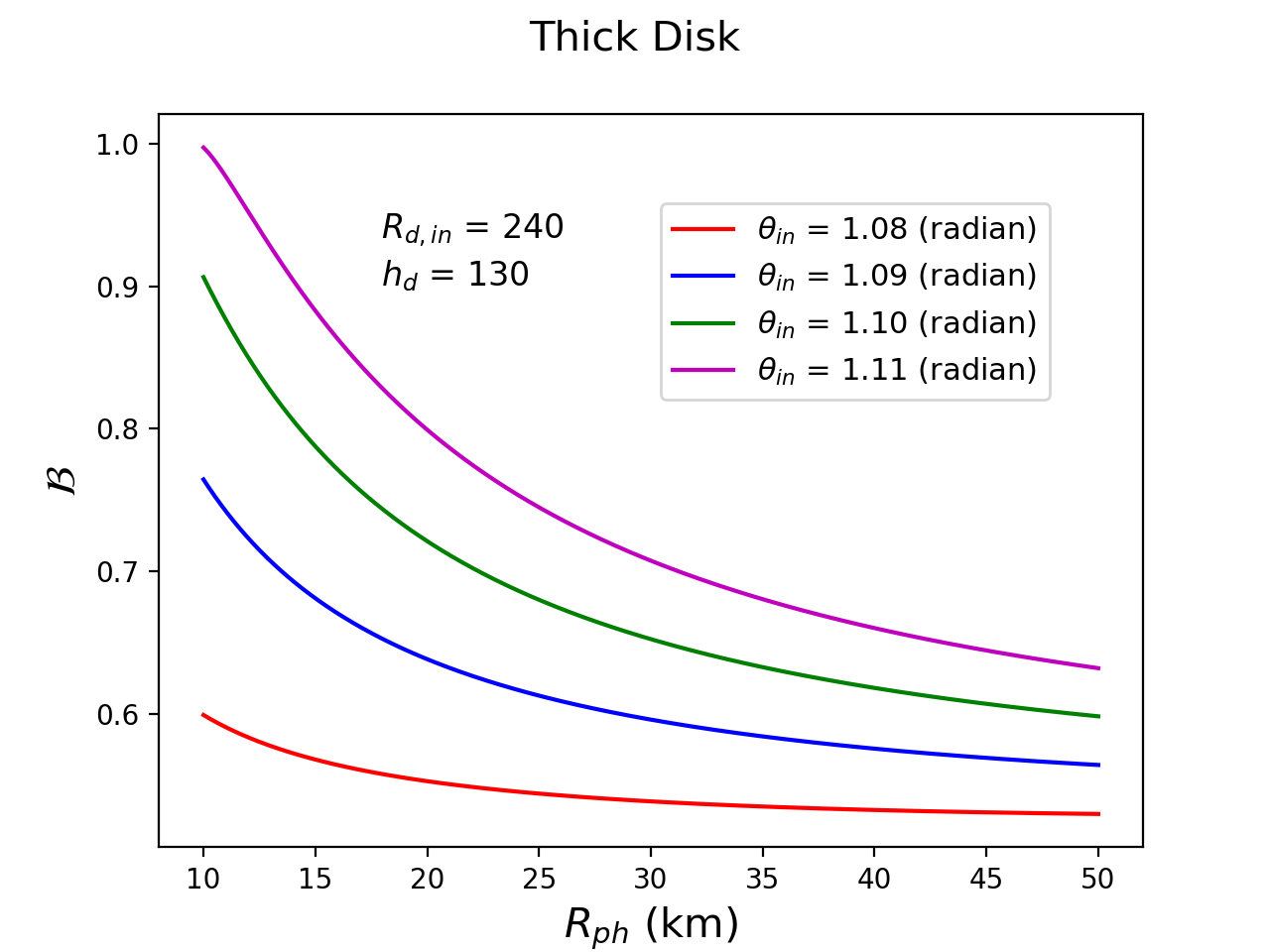}
      \label{alpha_vs_Rph_thickdisk}
   \end{subfigure}
   \caption{Blocking factor $\mathcal{B}$ for the thin-disk (top panels) and thick-disk (bottom panels) geometries shown in Fig.~1. Left panels: $\mathcal{B}$ vs.
   inclination angle $\theta_{\rm in}$ for fixed photospheric radius $R_{\rm ph}$. Right panels: $\mathcal{B}$ vs. $R_{\rm ph}$ for fixed inclination angle.
   In the thin-disk case, the relevant vertical scale is the disk half-thickness $h_d$, whereas in the thick-disk case the radial parameter shown is the inner disk radius $R_{d,\rm in}$,
   as defined in Fig.~1. The thin-disk model yields only modest blocking when $h_d < R_{\rm ph}$, while the thick-disk model shows rapid changes in $\mathcal{B}$
   as $\theta_{\rm in}$ approaches and exceeds the critical angle $\theta_{\rm cr}$.}
   \label{fig2_alpha}
\end{figure*}
The blocking factor serves as a key parameter in our reinterpretation of X-ray flux measurements from the neutron star in 4U 1746-37. This parameter provides
a quantitative approach to characterize the systematic reduction between the observed flux and the intrinsic emission from the neutron star surface.
By adopting this framework, we can account for the collective impact of processes that reduce the observed flux along the line of sight, without requiring
detailed modeling of each individual geometric component. This approach offers a practical means to reinterpret observational data while isolating the
net effect of flux reduction from the underlying complexity of the system.

The blocking factor $\mathcal{B}$ is defined as the fraction of the photosphere's projected area that is prevented from contributing to the
observed flux:
\begin{equation}
\label{eq1_blockingfactor}
\begin{aligned}
\mathcal{B} = \frac{\mbox{Blocked area of photosphere}}{\mbox{Total area of photosphere}}
\end{aligned}
\end{equation}

This definition inherently accounts for the observer's perspective and line of sight, making it a particularly useful tool for interpreting observations from
high-inclination systems where geometric effects significantly affect the observed properties. In theory, the blocking factor varies between 0 (no obscuration)
and 1 (complete obscuration), with intermediate values representing partial blocking scenarios that are commonly encountered in real astrophysical systems.

\subsection{Geometric dependencies and illustrative examples}
To demonstrate how the blocking factor responds to different geometric configurations, we utilize simplified disk models that capture the essential features
of accretion disk obscuration. Fig.~\ref{fig1_disks} illustrates two representative cases: a thin-disk configuration (left panel) and a thick-disk configuration
(right panel), both shown in cross-sectional view.
Our disk models are based on \citet{Shakura1973} and \citet{Narayan1994}; detailed specifications and parameters for these disk models are provided in the appendix.
For the illustrative calculations in Fig.~2, the thin-disk curves are constructed using the Shakura--Sunyaev thin-disk profile summarized in Table~A.1 (Case~1),
whereas the thick-disk curves are constructed using the advection-dominated accretion flow (ADAF) profile in Table~A.1 (Case~4).
For a given photospheric radius $R_{\rm ph}$ and inclination angle $\theta_{\rm in}$,
we evaluate the blocking factor numerically as
\[
\mathcal{B}(R_{\rm ph},\theta_{\rm in}) \equiv \frac{A_{\rm blocked}}{A_{\rm ph}},
\]
where $A_{\rm blocked}$ is the portion of the projected photosphere occulted by the adopted disk surface and $A_{\rm ph}$ is the total projected photospheric area. The curves shown in Fig.~2 are the resulting numerical evaluations of $\mathcal{B}$ as a function of $\theta_{\rm in}$ (left panels) or $R_{\rm ph}$ (right panels).
We emphasize that $\mathcal{B}$ is not directly observable in this framework; it is an effective parameter summarizing net line-of-sight obscuration.
The thin- and thick-disk models below are therefore illustrative and intended to demonstrate how disk structure and viewing geometry combine to determine the degree of
X-ray flux attenuation.

The thin-disk model represents a flattened, low-height structure characterized by a scale height $h_d$ that remains constant with radius. In this configuration,
the disk acts as a geometrically thin barrier that can partially obstruct the observer's view of the neutron star's photosphere, depending on the inclination angle
$\theta_{\rm in}$. For a spherical photosphere of radius $R_{\rm ph}$ centered at the disk's origin, blocking primarily occurs as the disk's inner and outer
edges lie in the observer's line of sight to the photospheric surface. At nearly edge-on viewing angles ($\theta_{\rm in} \approx {\pi}/2$),
complete obscuration ($\mathcal{B} = 1$) is not achieved when the disk height remains smaller than the photosphere radius ($h_d < R_{\rm ph}$).

The thick-disk model, by contrast, exhibits a scale height that increases with radius, characterized by a critical angle $\theta_{\rm cr}$ that defines the
disk's opening angle. This geometry can produce a different blocking behavior, particularly at high-inclination angles. 
As shown in Fig.~\ref{fig2_alpha} (bottom panels), the blocking factor for thick disks can exhibit sharp transitions as the inclination angle approaches and exceeds $\theta_{\rm cr}$.
In the thick-disk panels of Fig.~2, the quoted radial scale refers to the inner disk radius, $R_{d,\mathrm{in}}$, as defined in Fig.~1.
When $\theta_{\rm in} \approx \theta_{\rm cr}$, the observer's line of sight becomes tangent to the disk's inner edge, leading to rapid increases in the
blocking factor. For $\theta_{\rm in} > \theta_{\rm cr}$, the outer regions of the thick disk contribute primarily to the blocking, potentially achieving
high blocking factors ($\mathcal{B} > 0.8$) under reasonable geometric assumptions.

\subsection{Radius dependence and temporal evolution}
A fundamental aspect of the blocking factor is its dependence on the photosphere radius $R_{\rm ph}$, which varies significantly during the course of a
PRE burst. Fig.~\ref{fig2_alpha} (right panels) illustrates this relationship for both thin- and thick-disk configurations.
In general, larger photospheres experience less fractional blocking because a greater fraction of the photospheric surface extends beyond the region blocked
by the disk. This radius dependence has significant implications for interpreting burst observations, as the blocking factor evolves continuously throughout
the burst while the photosphere expands and contracts.

During the initial phases of a PRE burst, the photosphere may expand to several times the neutron star radius, substantially reducing the blocking factor
relative to its pre-burst value. As the burst progresses and the photosphere contracts back toward the neutron star surface during touchdown, the blocking
factor increases correspondingly. 
This temporal evolution of the blocking factor provides an explanation for the anomalously large peak-to-touchdown flux ratios observed in high-inclination systems like 4U 1746-37,
where these ratios exceed what can be accounted for by general relativistic effects alone. In particular, the observed peak-to-touchdown flux ratio provides an indirect constraint
on the difference between blocking at peak and touchdown, i.e.\ $\mathcal{B}(R_{\rm peak}) < \mathcal{B}(R_{\rm TD})$ (see Eq.~(\ref{eq7:fpeaktd}) in Sect.~\ref{sec3.4.2:photosphereevolution}).

\subsection{Practical implementation and observational implications}
The blocking factor's primary utility lies in its role as a summary parameter that quantifies the total observational impact of geometric obscuration
while remaining independent of the specific disk structure or precise viewing geometry. This approach acknowledges that the underlying physics of accretion disk structure is
complex and model-dependent, but that its observational consequence, namely a systematic reduction in the observed flux, can be characterized through a single parameter
that directly relates to the measurements.

For 4U 1746-37, the implementation of the blocking factor involves scaling both the observed touchdown flux $F_{\rm ob}$ and the effective emitting area
$A$ by the factor $(1-\mathcal{B})^{-1}$ to recover the intrinsic emission properties needed for mass and radius estimation. This scaling reflects the assumption
that the spectrum shape remains unchanged by geometric blocking, while the total flux is reduced proportionally to the fraction of the photosphere that remains
visible to the observer.

The blocking factor approach also provides a framework for understanding other observational features of high-inclination bursting sources, including the
correlation between dipping behavior in the persistent emission and sub-luminous burst properties. Systems that exhibit both characteristics are likely to represent
viewing geometries where significant blocking affects both the steady and burst emission, creating a consistent observational signature of high-inclination angles
combined with substantial disk thickness.

This framework emphasizes that the blocking factor, while informed by geometric considerations, fundamentally represents an observational parameter that encodes
the net effect of all processes that reduce the observed flux below its intrinsic value. By focusing on this observational perspective rather than attempting to
model each geometric detail, the blocking factor can provide a practical tool for addressing systematic observational biases in high-inclination LMXB systems. 

\section{Revised estimation of mass and radius}

\subsection{Analytical framework with blocking corrections}
The fundamental approach for determining neutron star masses and radii from PRE X-ray bursts relies on two key observational quantities measured:
the touchdown flux and the effective emitting area. However, in high-inclination systems like 4U 1746-37, geometric blocking can systematically reduce both quantities,
leading to underestimated intrinsic neutron star properties.

\subsubsection{Standard PRE equations}
The standard analysis of PRE bursts employs two fundamental equations that connect observable quantities to neutron star properties \citep{vanParadijs1987,Damen1990}.
The touchdown flux $F_{\rm TD}$, representing the Eddington-limited emission when the photosphere contracts back to the neutron star surface, is given by:

\begin{equation}
\label{eq2_touchdownflux}
\begin{aligned}
F_{\rm TD} = \frac{cGM_{\rm NS}}{\kappa D^2} \left(1 - \frac{2GM_{\rm NS}}{R_{\rm NS}c^2}\right)^{1/2}
\end{aligned}
\end{equation}

where $c$ is the speed of light, $G$ is the gravitational constant, $M_{\rm NS}$ and $R_{\rm NS}$ are respectively the neutron star mass and radius,
$\kappa$ is the opacity, and $D$ is the distance to the source. 
The general relativistic correction factor accounts for the gravitational redshift of the radiation emitted from the neutron star surface.

The effective emitting area $A$, derived from the observed bolometric flux and blackbody temperature during the burst cooling phase, is expressed as

\begin{equation}
\label{eq3_effectiveemittingarea}
\begin{aligned}
A = \frac{F_\infty}{\sigma_{\rm {SB}}T_{\rm {bb}}^4} = f_{\rm c}^{-4} \frac{R_{\rm NS}^2}{D^2} \left(1 - \frac{2GM_{\rm NS}}{R_{\rm NS}c^2}\right)^{-1}
\end{aligned}
\end{equation}

where $F_\infty$ is the observed bolometric flux, $\sigma_{\rm {SB}}$ is the Stefan-Boltzmann constant, $T_{\rm {bb}}$ is the blackbody temperature, and
$f_{\rm c}$ is the color-correction factor that accounts for spectral hardening in the neutron star atmosphere. 

\subsubsection{Blocking factor implementation}
In systems with significant geometric obscuration, the observed quantities $F_{\rm {TD,ob}}$ and $A_{\rm {ob}}$ represent only the unobscured components
of the true emission. The blocking factor $\mathcal{B}$, defined as the fraction of the photosphere area obscured from view, provides the correction mechanism
so that the relationship between observed and intrinsic quantities becomes

\begin{equation}
\label{eq4_observedflux}
\begin{aligned}
F_{\rm {ob}} = (1 - \mathcal{B}) F_{\rm {tr}}
\end{aligned}
\end{equation}

where $F_{\rm {tr}}$ represents the true flux that would be observed in the absence of blocking.

\subsubsection{Modified PRE equations}
Incorporating the blocking factor into the standard PRE framework, we modify Eqs.~(\ref{eq2_touchdownflux}) and (\ref{eq3_effectiveemittingarea}).
During the touchdown phase, the photosphere typically contracts to approximately the neutron star radius, such that
$\mathcal{B}(R_{\rm {TD}}) \approx \mathcal{B}(R_{\rm {NS}})$. The corrected touchdown flux equation becomes:

\begin{equation}
\label{eq5_truetouchdownflux}
\begin{aligned}
F_{\rm {TD,tr}} = \frac{F_{\rm {TD,ob}}}{1 - \mathcal{B}(R_{\rm {NS}})}
= \frac{cGM_{\rm {NS}}}{\kappa D^2} \left(1 - \frac{2GM_{\rm {NS}}}{c^2R_{\rm {NS}}}\right)^{1/2}
\end{aligned}
\end{equation}

Similarly, the corrected effective area is:

\begin{equation}
\label{eq6_trueeffectivearea}
\begin{aligned}
A_{\rm {tr}} = \frac{A_{\rm {ob}}}{1 - \mathcal{B}(R_{\rm {NS}})}
= f_{\rm c}^{-4} \frac{R_{\rm {NS}}^2}{D^2} \left(1 - \frac{2GM_{\rm {NS}}}{c^2R_{\rm {NS}}}\right)^{-1}
\end{aligned}
\end{equation}

\subsection{Observational parameters and implementation}
\label{sec3.2:obs_param}
\subsubsection{Input parameters for 4U 1746-37}
\begin{table}[ht]
  \centering
  \caption{Input parameters adopted for 4U 1746-37.}
  \begin{tabular}{ll}
    \hline
    Parameter & Value \\
    \hline
    Touchdown flux $F_{\mathrm {TD,ob}}$ & $(2.69 \pm 0.57) \times 10^{-9}$ erg\,s$^{-1}$\,cm$^{-2}$ \\
    Effective area $A_\mathrm {ob}$ & $10.9 \pm 4.4$ (km/10 kpc)$^2$ \\
    Distance $D$ & $11.05 \pm 0.85$ kpc \\
    Color-correction factor $f_{\mathrm c}$ & $1.35 \pm 0.05$ \\
    Hydrogen mass fraction $X$ & $0.35 \pm 0.35$ \\
    Opacity $\kappa$ & $0.2(1+X)$ cm$^2$\,g$^{-1}$ \\
    \hline
  \end{tabular}
  \label{table1:input4u1746}
\end{table}
Following the analysis of \citet{Li2015}, we adopt the same observational parameters and their uncertainties for consistency:
The choice of uniform distributions for $f_{\rm c} \in [1.3, 1.4]$ and $X \in [0, 0.7]$ reflects the current uncertainties in atmospheric modeling
and composition determination, respectively. 
While we acknowledge that $f_{\rm c}$ values may vary during PRE evolution and can potentially reach higher values ($f_{\rm c} \approx 2.0$) under certain conditions,
we maintain this specific range to ensure that any deviations in our results arise solely from the introduction of the blocking factor.
The implications of these broader parameter variations are discussed in detail in Sect.~\ref{sec:colorcorrection}.
%% Figure 3
\begin{figure}[!ht]
\centering
\includegraphics[width=\hsize]{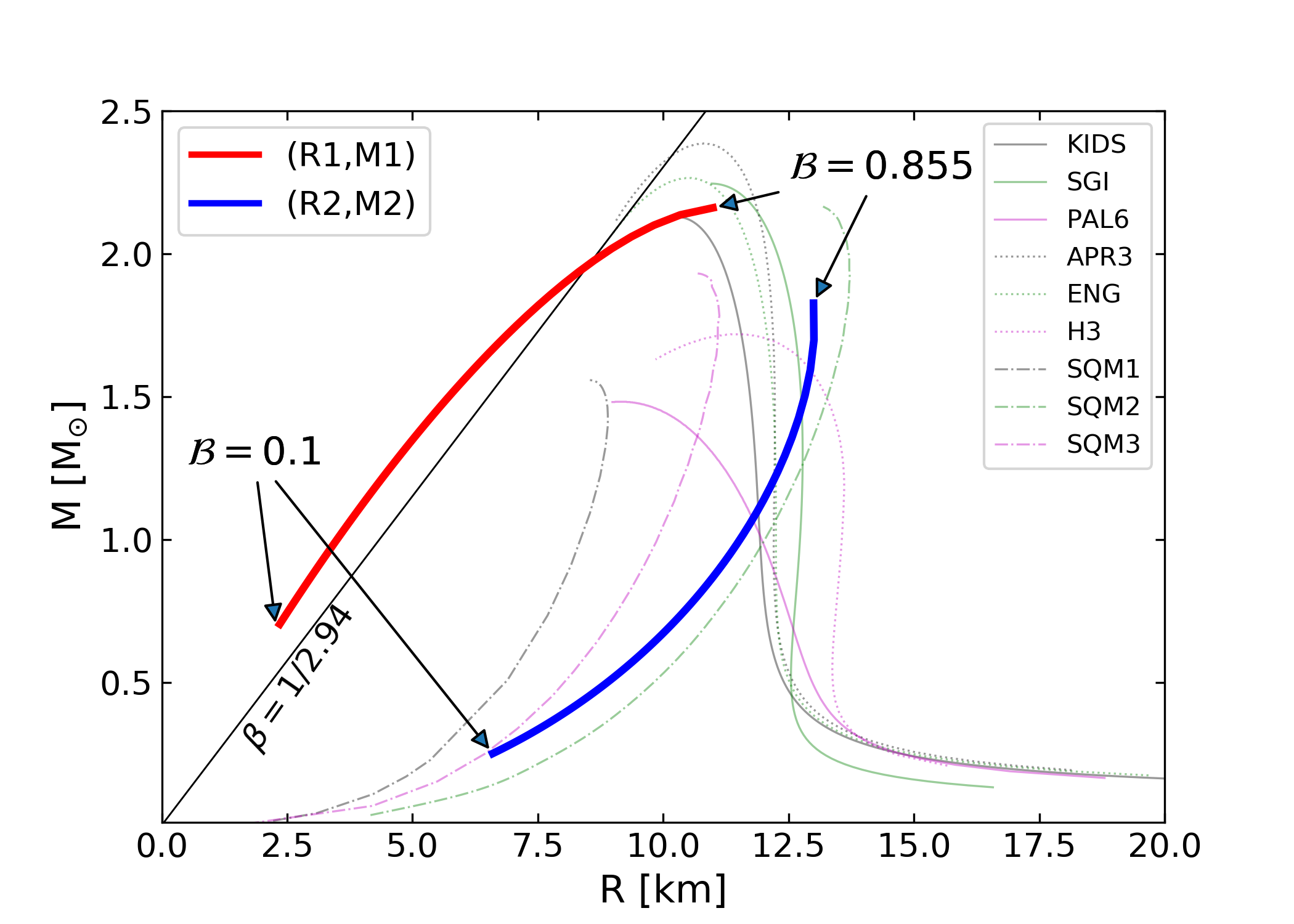}
\caption{Mass--radius solutions for the neutron star in 4U 1746-37 as a function of blocking factor $\mathcal{B}$. The blue and red curves represent two analytical solution branches $(R_1,M_1)$ and $(R_2,M_2)$ obtained from solving the modified PRE equations for blocking factors ranging from $\mathcal{B} = 0.1$ to $\mathcal{B} = 0.855$. The black diagonal line shows the general relativistic causality limit $\beta \equiv \frac{GM}{Rc^2} = 1/2.94$. The $(R_1,M_1)$ branch violates this constraint at small $\mathcal{B}$ values, while $(R_2,M_2)$ remains physically viable across a broader parameter range. Overlaid are nine equations of state (EoS) models: six normal nuclear matter models (KIDS, SGI, PAL6, APR3, ENG, H3)
and three quark matter models (SQM1, SQM2, SQM3). Significant blocking $\mathcal{B} \gtrsim 0.8$ can reconcile neutron star parameters with canonical values predicted by
normal nuclear matter equations of state.}
\label{fig3_line}
\end{figure}

\subsubsection{Solution methodology}
For moderate blocking factors $\mathcal{B} \lesssim 0.85$, the system of modified PRE equations can be reduced to a single quadratic equation for the
neutron star compactness, $\beta \equiv GM_{\mathrm {NS}}/R_{\mathrm {NS}}c^2$. By combining Eqs.~(5) and (6) to eliminate the mass and radius, we obtain:

\begin{equation}
    \beta(1-2\beta) = \alpha_{\rm ob} (1-\mathcal{B})^{-1/2}
    \label{eq:compactness_quadratic}
\end{equation}

\noindent where $\alpha_{\rm ob}$ groups the observational quantities:

\begin{equation}
    \alpha_{\rm ob} \equiv \frac{\kappa D F_{\mathrm {TD,ob}}}{c^3 f_c^2 \sqrt{A_{\mathrm{ob}}}}
    \label{eq:alpha_obs}
\end{equation}

\noindent Solving for $\beta$ yields two distinct roots:

\begin{equation}
    \beta_{\pm} = \frac{1 \pm \sqrt{1 - 8\alpha_{\rm ob}(1-\mathcal{B})^{-1/2}}}{4}
    \label{eq9:beta_roots}
\end{equation}

These two roots, $\beta_{-}$ and $\beta_{+}$, correspond directly to the two solution branches $(M_1, R_1)$ and $(M_2, R_2)$ shown in Fig.~\ref{fig3_line}.
The lower root $\beta_{-}$ (the $(M_2, R_2)$ branch) represents the physically stable solution that typically satisfies the general relativistic causality limit ($\beta < 1/2.94$).

\begin{description}
    \item[The $(M_1, R_1)$ branch:] (red line in Fig.~\ref{fig3_line}); this branch produces solutions that violate general relativistic causality constraints
    for small blocking factors ($\mathcal{B} \lesssim 0.6$). These solutions correspond to where the compactness parameter $\beta \equiv \frac{GM}{Rc^2}$ exceeds
    the causality limit, rendering them unphysical.
    \item[The $(M_2, R_2)$ branch:] (blue line in Fig.~\ref{fig3_line}); this branch provides physically viable solutions across a broader range of blocking
    factors. These solutions remain below the causality limit and yield neutron star parameters that approach canonical values as $\mathcal{B}$ increases.
\end{description}
The general relativistic causality limit is defined by the compactness parameter $\beta \equiv \frac{GM}{Rc^2} = {1}/{2.94}$. This boundary represents the
theoretical maximum compactness for stable neutron star configurations. Solutions exceeding this limit are excluded from our analysis as they violate fundamental
physical constraints.

As the blocking factor $\mathcal{B}$ increases, the term $(1-\mathcal{B})^{-1/2}$ grows larger, causing the quantity under the square root (the discriminant) to decrease.
If $\mathcal{B}$ becomes sufficiently large, the discriminant may become negative, indicating that no physical solution exists for that specific set of observational parameters.
This mathematical constraint explains why analytical solutions become unstable or non-existent at very high blocking factors, necessitating the statistical Monte Carlo approach
described in Sect.~\ref{sec3.3.3:mchigh} to explore the parameter space where valid solutions persist.

More generally, independent of whether the solutions are obtained analytically or via Monte Carlo, the touchdown method provides two algebraic constraints based on
$F_{\rm TD}$ and $A$. The present framework does not permit a joint statistical inference of $M$, $R$, and $\mathcal{B}$ from the burst data alone.
In this work, $\mathcal{B}$ is therefore treated as an externally specified diagnostic parameter rather than a fitted quantity. Accordingly, we do not perform a likelihood or
evidence comparison among different $\mathcal{B}$ values; instead, we examine how the inferred mass--radius solutions change across an assumed range of
$\mathcal{B}$ and how the fraction of physically admissible Monte Carlo realizations varies.

\begin{table*}[!ht]
\caption{Mass and radius estimates from propagation of error analysis}
\label{table2:err}
\centering
\begin{tabular}{ccccccccc}
\toprule\toprule
$(1-\mathcal{B})^{-1}$ & $M_1$ & $\sigma_{M_1}$ & $R_1$ & $\sigma_{R_1}$ & $M_2$ & $\sigma_{M_2}$ & $R_2$ & $\sigma_{R_2}$ \\
(scaling factor) & ($\rm M_{\sun}$) & ($\rm M_{\sun}$) & (km) & (km) & ($\rm M_{\sun}$) & ($\rm M_{\sun}$) & (km) & (km) \\
\midrule
2 & $\cdot\cdot\cdot$ & $\cdot\cdot\cdot$ & $\cdot\cdot\cdot$ & $\cdot\cdot\cdot$ & 0.463 & 0.146 & 8.63 & 2.18 \\
3 & $\cdot\cdot\cdot$ & $\cdot\cdot\cdot$ & $\cdot\cdot\cdot$ & $\cdot\cdot\cdot$ & 0.714 & 0.233 & 10.3 & 2.74 \\
4 & $\cdot\cdot\cdot$ & $\cdot\cdot\cdot$ & $\cdot\cdot\cdot$ & $\cdot\cdot\cdot$ & 0.982 & 0.352 & 11.5 & 3.32 \\
5 & $\cdot\cdot\cdot$ & $\cdot\cdot\cdot$ & $\cdot\cdot\cdot$ & $\cdot\cdot\cdot$ & 1.27 & 0.502 & 12.4 & 4.06 \\
6 & 2.12 & 1.08 & 9.80 & 4.13 & 1.59 & 0.689 & 13.0 & 5.45 \\
\bottomrule
\end{tabular}
\tablefoot{Blank entries indicate unphysical solutions violating the general relativistic causality constraint $\frac{GM}{Rc^{2}} < \frac{1}{2.94}$.}
\end{table*}

\subsection{Results for different blocking scenarios}
\subsubsection{Moderate- to high-blocking cases}
Table \ref{table2:err} presents the systematic evolution of neutron star mass and radius as functions of the scaling factor $(1 - \mathcal{B})^{-1}$.
To determine the associated uncertainties, we employ an analytical propagation-of-errors technique following the methodology established by \citet{Steiner2010}.
This approach derives $\sigma_{M_{\rm{NS}}}$ and $\sigma_{R_{\rm{NS}}}$ directly from the observational uncertainties in $F_{\rm{TD}}$, $A$, $D$, $f_{\rm{c}}$, and $X$
while treating the scaling factor itself as a fixed parameter.

\begin{description}
    \item[Scaling factor = 5:]
      $M_2 = 1.27 \pm 0.50 ~\rm{M_{\sun}}$, $R_2 = 12.4 \pm 4.1 ~\rm{km}$\\
    \item[Scaling factor = 6:]
    \begin{minipage}{\linewidth}
        $M_1 = 2.12 \pm 1.08 ~\rm{M_{\sun}}$, $R_1 = 9.80 \pm 4.13 ~\rm{km}$ \\
        $M_2 = 1.59 \pm 0.69 ~\rm{M_{\sun}}$, $R_2 = 13.0 \pm 5.45 ~\rm{km}$
    \end{minipage}
\end{description}
These results demonstrate a crucial transition: with blocking factors of $\mathcal{B} \sim 0.8-0.85$, the derived neutron star parameters approach canonical values
that are consistent with normal nuclear matter equations of state. 

As the scaling factor increases from 2 to 6, both the central values and uncertainties of the mass and radius estimates increase systematically.
The results indicate that uncertainties remain relatively large even with constant blocking factors, reflecting the inherent observational limits of the input parameters.
Furthermore, the solution branch ($M_1$, $R_1$) becomes physically viable only at these higher scaling factors where the general relativistic causality constraint is satisfied.

\begin{figure*}[!ht]
\sidecaption
\includegraphics[width=12cm]{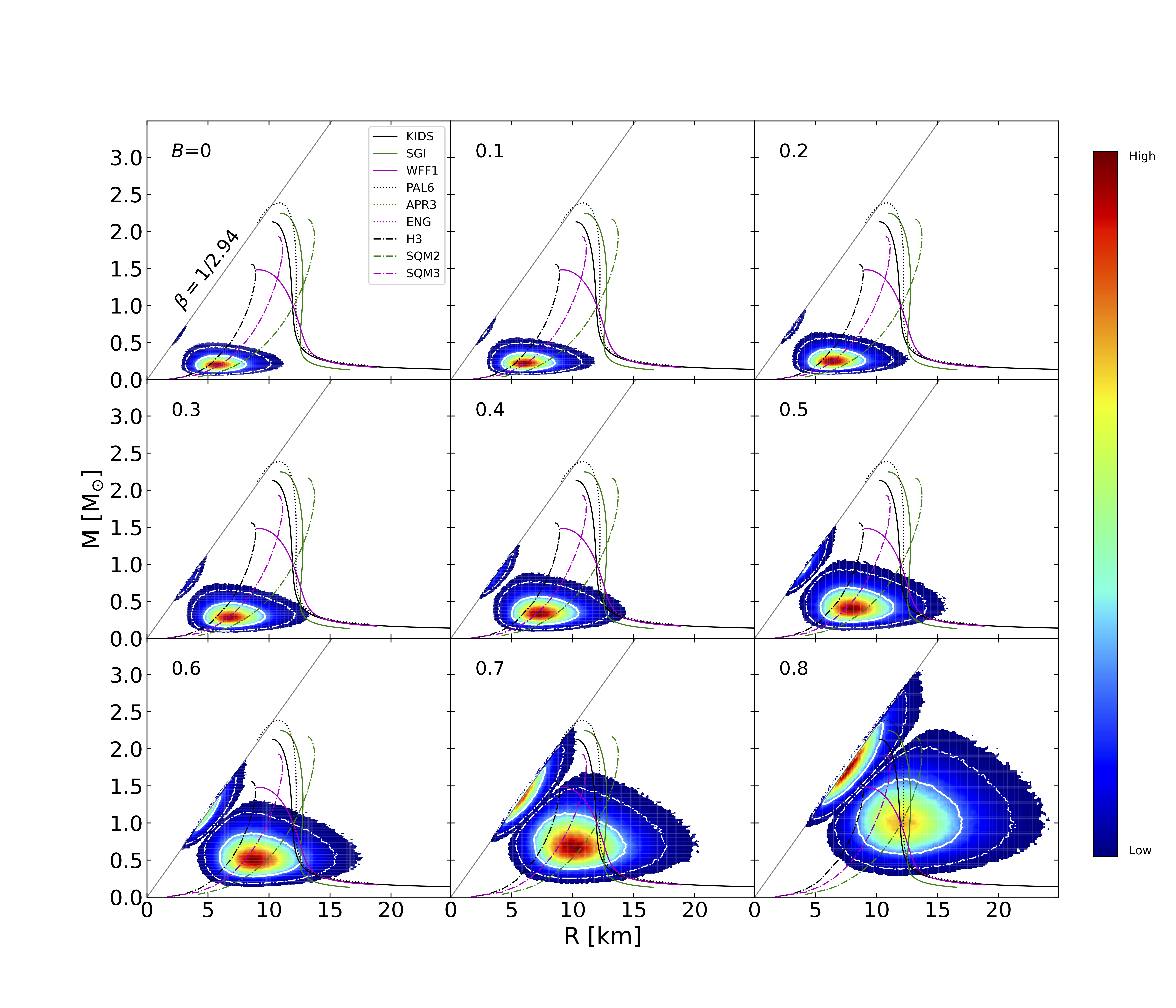}
\caption{Monte Carlo mass--radius constraints for 4U 1746$-$37 over a sampled grid of fixed blocking factors, $\mathcal{B}=0, 0.1, 0.2, \ldots, 0.8$. Each panel shows the distribution of accepted solutions in the $M$--$R$ plane for the indicated value of $\mathcal{B}$. As the assumed blocking factor increases, the allowed region shifts systematically from the unusually compact, low-mass regime toward higher masses and larger radii. The white solid and dot-dashed contours enclose $68\%$ and $95\%$ of the accepted realizations, respectively. The gray diagonal line marks the causality limit, $\beta \equiv GM/(Rc^{2}) = 1/2.94$. Overlaid are representative equations of state for nucleonic matter (KIDS, SGI, PAL6, APR3, ENG, H3) and strange-quark matter (SQM1, SQM2, SQM3).}
\label{fig4:bscan1}
\end{figure*}

\begin{figure*}[!ht]
\sidecaption
\includegraphics[width=12cm]{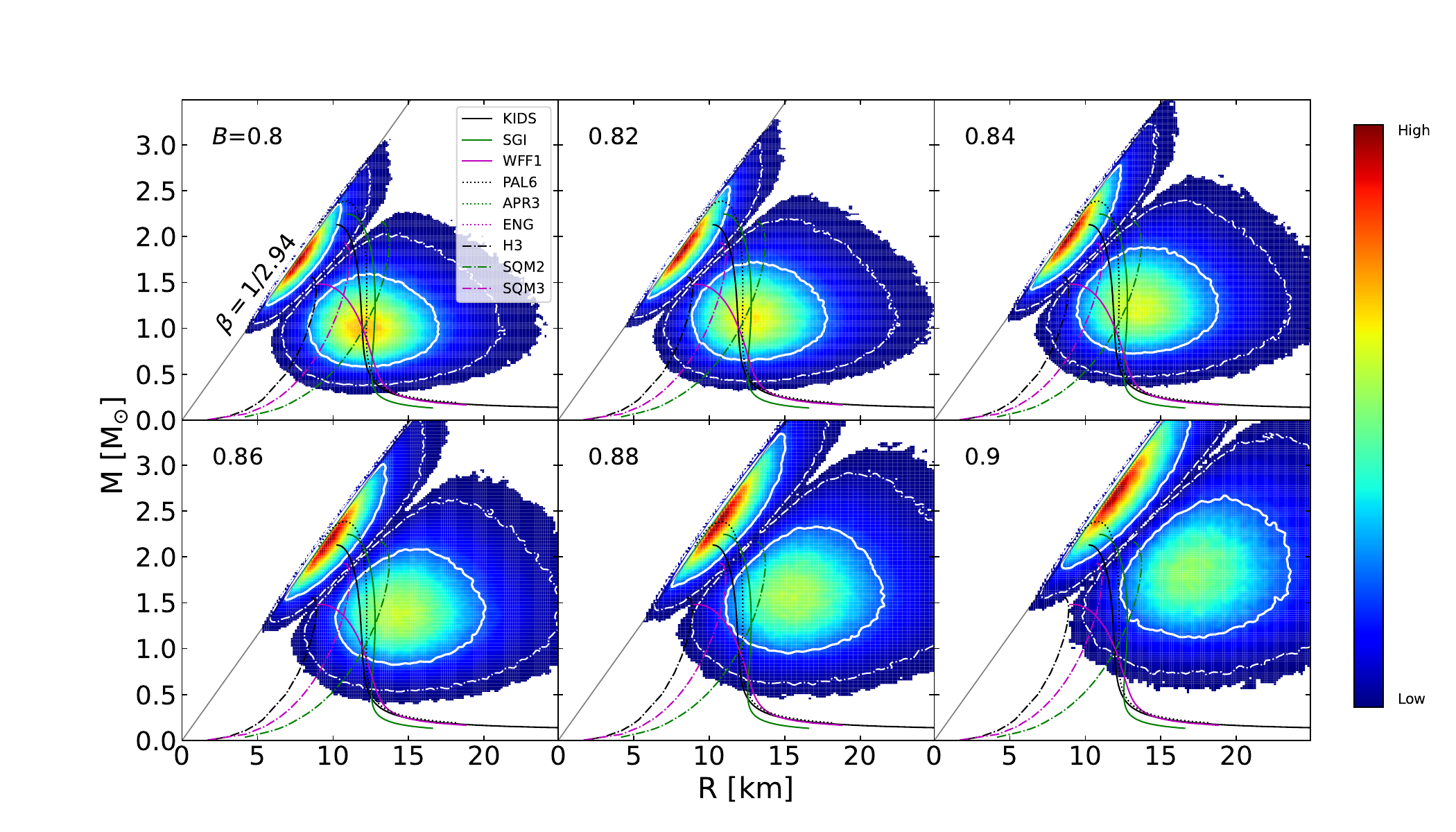}
\caption{Zoomed Monte Carlo mass--radius constraints for 4U 1746$-$37 in the high-blocking regime, $\mathcal{B}=0.8, 0.82, 0.84, 0.86, 0.88,$ and $0.9$. This finer sampling highlights the rapid evolution of the allowed solution space as $\mathcal{B}$ approaches extreme values. In this regime, the surviving solutions extend into the canonical neutron star range and overlap more substantially with several standard nucleonic equations of state. The white solid and dot-dashed contours enclose $68\%$ and $95\%$ of the accepted realizations, respectively, and the gray diagonal line denotes the causality limit, $\beta = 1/2.94$.}
\label{fig5:bscan2}
\end{figure*}

\begin{table*}[!t]
\caption{Monte Carlo mass--radius constraints for representative fixed blocking factors}
\label{table3:MC}
\centering
\begin{tabular}{lcccccccc}
\toprule\toprule
& \multicolumn{2}{c}{\makecell{No blocking ($\mathcal{B}$\,=\,0)}} &
  \multicolumn{2}{c}{\makecell{Intermediate blocking ($\mathcal{B}$\,=\,0.5)}} &
  \multicolumn{2}{c}{\makecell{Intermediate blocking ($\mathcal{B}$\,=\,0.7)}} &
  \multicolumn{2}{c}{\makecell{High blocking ($\mathcal{B}$\,=\,0.9)}} \\
\cmidrule(lr){2-3}\cmidrule(lr){4-5}\cmidrule(lr){6-7}\cmidrule(lr){8-9}
& $\beta_{-}$ sol. & $\beta_{+}$ sol. &
  $\beta_{-}$ sol. & $\beta_{+}$ sol. &
  $\beta_{-}$ sol. & $\beta_{+}$ sol. &
  $\beta_{-}$ sol. & $\beta_{+}$ sol. \\
\midrule
$M_{\mathrm{NS}}^{a}$ & 0.207 & \dots$^{b}$ & 0.405 & \dots$^{b}$ & 0.647 & \dots$^{b}$ & 1.86 & 2.61 \\
$R_{\mathrm{NS}}^{c}$ & 5.69  & \dots$^{b}$ & 7.85  & \dots$^{b}$ & 10.35 & \dots$^{b}$ & 17.3 & 11.6 \\
Sol.$^{d}$            & 99.9  & 0.4         & 97.2  & 4.4         & 86.3  & 11.6        & 29.6 & 11.8 \\
Causality$^{e}$       & 0.0   & 99.5        & 0.0   & 92.8        & 0.0   & 74.7        & 0.0  & 17.8 \\
Unphys.$^{f}$         & 0.1   & 0.1         & 2.8   & 2.8         & 13.7  & 13.7        & 70.4 & 70.4 \\
\bottomrule
\end{tabular}
\vspace{1ex}
\vfill
\raggedright
$^{a}$~Most probable value of neutron star mass (in solar masses).\\
$^{b}$~Not reported because the solution acceptance fraction is too low to yield a statistically significant estimate.\\
$^{c}$~Most probable value of neutron star radius (in kilometers).\\
$^{d}$~Acceptance fraction in the entire MC sampling, defined for each solution branch as the fraction of realizations that yield a real
algebraic solution and satisfy the causality condition.\\
$^{e}$~Fraction in the entire MC sampling that violates the causality condition.\\
$^{f}$~Fraction in the entire MC sampling for which no algebraic solution exists.
\tablefoot{Each sampled blocking factor is based on $2\times10^{6}$ Monte Carlo realizations. A single MC sample can produce two solutions.
The $\beta_{+}$ solution lies closer to the causality limit $(\beta=1/2.94)$ than the $\beta_{-}$ solution.}
\end{table*}

\subsubsection{Monte Carlo scan over fixed blocking factors}
\label{sec3.3.2:bscan}
To clarify whether the burst data require high blocking factors or merely allow them, we performed a Monte Carlo (MC) scan over a grid of fixed blocking factors,
$\mathcal{B}=0, 0.1, 0.2, \dots, 0.9$. For each assumed value of $\mathcal{B}$, we generated $2\times10^{6}$ realizations by sampling the observed parameter distributions
and solving the modified PRE equations for the corresponding mass--radius solutions. Because the touchdown method provides only two algebraic constraints based on
$F_{\rm TD}$ and $A$, this procedure does not constitute a statistical fit for $\mathcal{B}$ and should not be interpreted as yielding a likelihood or posterior preference for
any particular blocking factor. Instead, it serves as a diagnostic test of how the inferred solution topology and the fraction of physically admissible realizations change across
an assumed range of $\mathcal{B}$.

Figs.~\ref{fig4:bscan1} and \ref{fig5:bscan2} show the resulting evolution of the allowed $M$--$R$ regions as $\mathcal{B}$ increases. At low blocking factors,
the physically admissible solutions are dominated by the $\beta_{-}$ branch and remain concentrated in the unusually low-mass compact regime reported by \citet{Li2015}.
In contrast, the $\beta_{+}$ branch is strongly suppressed at low $\mathcal{B}$ because most realizations violate the causality condition $\beta \equiv GM/(Rc^{2}) < 1/2.94$.
As the assumed blocking factor increases, the allowed contours broaden and shift systematically toward larger masses and radii. This trend demonstrates that geometric blocking
can move the inferred neutron star parameters away from the compact, exotic-matter-like regime and toward values more consistent with canonical neutron stars.

To provide a compact numerical summary of this behavior, Table~\ref{table3:MC} lists representative sampled cases at $\mathcal{B}=0$, $0.5$, $0.7$, and $0.9$.
For each branch, we report the fraction of MC realizations that yield real algebraic solutions, the fraction excluded by the causality condition, and the fraction for which no real solution
exists. These quantities are diagnostic indicators of physical admissibility within the assumed MC sampling, rather than fit statistics for $\mathcal{B}$. The scan shows that
the $\beta_{-}$ branch remains the dominant physically admissible branch over most of the explored range, while the contribution from the $\beta_{+}$ branch grows only at
high blocking factors. Thus, the present analysis shows that large blocking factors are allowed and can reconcile the burst observables with canonical neutron star parameters,
but it does not establish that such values are statistically preferred by the data.

\subsubsection{High-blocking regime}
\label{sec3.3.3:mchigh}
For blocking factors approaching the extreme regime, $\mathcal{B} \gtrsim 0.8$--$0.9$, the analytical solutions become increasingly fragile because the factor
$(1-\mathcal{B})^{-1}$ grows rapidly and the discriminant in Eq.~(\ref{eq9:beta_roots}) approaches zero or becomes negative for a substantial fraction of the
sampled parameter combinations. In this regime, the Monte Carlo approach provides a more robust way to visualize the surviving solution space and to assess the distribution
of physically admissible solutions.

As shown by the high $\mathcal{B}$ panels in Figs.~4 and 5, the allowed mass--radius contours shift substantially toward larger masses and radii as the assumed
blocking factor increases. In contrast to the low-blocking cases, where the physically admissible solutions remain concentrated in the unusually compact regime reported
by \citet{Li2015}, the high-blocking regime extends into the canonical neutron star range and overlaps more substantially with several standard nucleonic equations of state.

The Monte Carlo results also clarify the distinct roles of the two solution branches. The $\beta_{-}$ branch remains the dominant physically admissible branch over most of the
explored range, while the $\beta_{+}$ branch lies closer to the causality boundary and is therefore more strongly restricted. At sufficiently high blocking factors,
however, both branches can survive in part of the sampled parameter space and occupy distinct regions in the $M$--$R$ plane.

Most importantly, these high-blocking results should not be interpreted as showing that the data favor $\mathcal{B} \sim 0.9$. Rather, together with the broader
fixed-$\mathcal{B}$ scan in Sect.~\ref{sec3.3.2:bscan}, they show that sufficiently strong blocking provides a viable alternative interpretation of the burst data. When the reported
statistical uncertainties in $F_{\rm TD,ob}$, $A_{\rm ob}$, $D$, $f_c$, and $X$ are propagated through the modified PRE equations, the resulting $M$--$R$ constraints can shift
from the anomalously compact regime toward values compatible with canonical neutron stars, while remaining subject to the additional systematic uncertainties discussed in Sect.~\ref{sec:4.3}.

\subsection{Peak-to-touchdown flux ratio analysis}
\subsubsection{Blocking factor explanation}
The observed peak-to-touchdown flux ratio in 4U 1746-37, $f_{\rm {peak/TD}} \approx 2.0$, exceeds the maximum ratio ($\sim 1.3$) that can be explained by
general relativistic effects alone, creating a fundamental discrepancy between observations and theoretical predictions.

The radius-dependent nature of the blocking factor provides a natural explanation for this anomalously large ratio. During PRE evolution, the photosphere can
expand from the neutron surface to several times the neutron star radius $R_{\rm {NS}}$, at peak luminosity, then contracts back during touchdown.
As the photosphere expands, the blocking factor decreases because a larger fraction of the photospheric surface extends beyond the geometrically obscured region.

This creates a systematic difference between the blocking factors: $\mathcal{B}(R_{\rm{peak}}) \ll \mathcal{B}(R_{\rm{TD}})$. The observed peak-to-touchdown
flux ratio then reflects this differential blocking:

\begin{equation}
\label{eq7:fpeaktd}
\begin{aligned}
f_{\rm {peak/TD}} = \frac{1 - \mathcal{B}(R_{\rm {peak}})}{1 - \mathcal{B}(R_{\rm {TD}})}
                  \sqrt{\frac{1 - \frac{2GM_{\rm {NS}}}{c^2R_{\rm {peak}}}}{1 - \frac{2GM_{\rm {NS}}}{c^2R_{\rm {TD}}}}}
\end{aligned}
\end{equation}

The first term captures the geometric blocking effect that naturally accounts for the observed ratio of $\sim 2.0$, while the second term provides general
relativistic corrections at different photospheric radii.

\subsubsection{Photosphere evolution constraints}
\label{sec3.4.2:photosphereevolution}
To quantify the impact of radius-dependent blocking, we estimate characteristic parameters using the thick-disk model at a representative inclination angle $\theta_{\rm in} \approx 1.10$ 
rad. We take the well-constrained neutron star parameters from our blocking-factor analysis: mass $M_{\rm NS} = 1.59\,M_\odot$ and touchdown radius $R_{\rm TD}$ equal to the neutron star radius $R_{\rm NS} = 13$ km, along with the blocking factor at touchdown $\mathcal{B}(R_{\rm TD}) \approx 0.83$. 
This $\mathcal{B}(R)$ corresponds to the numerically evaluated thick-disk profile shown by the green curve in the bottom-right panel of Fig.~2, constructed from the
Case~4 (ADAF) disk geometry in Table~A.1 at fixed $\theta_{\rm in}\approx 1.10~{\rm rad}$.
This choice of $\theta_{\rm in}$ and the associated $\mathcal{B}(R)$ curve is representative and is used to illustrate the implications of radius-dependent blocking;
other $\mathcal{B}(R)$ forms consistent with the geometry could also be adopted.

With the observed $f_{\rm peak/TD}$ ratio, neutron star mass, touchdown radius, and blocking at touchdown as inputs, and adopting the representative blocking profile
$\mathcal{B}(R)$ from Fig.~\ref{fig2_alpha}, we solve Eq.~(\ref{eq7:fpeaktd}) numerically (root-finding) for the single unknown $R_{\rm peak}$.
This is necessary because $\mathcal{B}(R_{\rm peak})$ depends on the unknown $R_{\rm peak}$ through the adopted $\mathcal{B}(R)$ profile, so the resulting equation
does not admit a closed-form solution for a general $\mathcal{B}(R)$. The corresponding $\mathcal{B}(R_{\rm peak})$ is then obtained by evaluating $\mathcal{B}(R)$ at that radius.
The solution yields a peak radius $R_{\rm peak} \approx 23.5$ km and
a correspondingly smaller blocking factor $\mathcal{B}(R_{\rm peak}) \approx 0.69$, reflecting the photosphere’s expansion beyond the geometrically obscured inner disk region
during the peak-luminosity phase.

The contraction of the photosphere from $R_{\rm peak}$ to $R_{\rm TD}$ corresponds to a net decrease in visible emitting area and an increase in blocking factor as the burst decays. The radial contraction speed is approximately $5.3\ {\rm km\ s^{-1}}$, which is much lower than the typical isothermal sound speed under these conditions, $\sim 560\ {\rm km\ s^{-1}}$. This indicates that the contraction phase is highly subsonic, as expected for a radiatively regulated photospheric evolution.

We emphasize that this calculation is intended only as an illustrative plausibility check. The inferred $R_{\rm peak}$ depends most strongly on the adopted radius-dependent
blocking profile $\mathcal{B}(R)$ (i.e., the viewing geometry, parameterized here by $\theta_{\rm in}$) and on the assumed touchdown radius $R_{\rm TD}$, whereas the dependence
on $M_{\rm NS}$ enters only through the relatively weak GR correction in Eq.~(10). Varying $\theta_{\rm in}$ (and thus $\mathcal{B}(R)$) or $R_{\rm TD}$ therefore shifts the
inferred $R_{\rm peak}$ at the tens-of-percent level, but the qualitative conclusion remains unchanged: for geometries capable of producing the large differential blocking required by
the observed $f_{\rm peak/TD}$, the contraction speed remains orders of magnitude below the sound speed,
consistent with a highly subsonic, radiatively regulated photospheric evolution.

\begin{table}[ht]
\centering
\caption{Equation of state models used for comparison with neutron star mass--radius constraints}
\begin{tabular}{ll}
\hline
Model & Reference \\
\hline
{Normal nuclear matter} & \\
\quad KIDS & Papakonstantinou et al. (2018) \\
\quad SGI & Li (1991) \\
\quad PAL6 & Prakash et al. (1988) \\
\quad APR3 & Akmal et al. (1998) \\
\quad ENG & Engvik et al. (1996) \\
\quad H3 (with hyperons) & Lackey et al. (2006) \\
{Quark matter} & \\
\quad SQM1 & Prakash et al. (1995) \\
\quad SQM2 & Prakash et al. (1995) \\
\quad SQM3 & Prakash et al. (1995) \\
\hline
\end{tabular}
\label{table4:eosmodels}
\end{table}

\subsection{Alternative scenarios: Touchdown away from surface}
\subsubsection{Extended touchdown analysis}
Previous studies have suggested that touchdown may not occur precisely at the neutron star surface, but instead at a larger radius,
$R_{\rm TD} > R_{\rm NS}$ \citep{Steiner2010,Kim2021}.
In this case, the effective-emitting-area relation remains tied to the neutron star surface, whereas the touchdown-flux relation must be evaluated at the touchdown radius.
We therefore solve the following modified PRE equations:
\begin{align*}
A_{\rm tr} &=
\frac{A_{\rm ob}}{1-\mathcal{B}(R_{\rm NS})}
= f_c^{-4}\frac{R_{\rm NS}^2}{D^2}
\left(1-\frac{2GM_{\rm NS}}{c^2R_{\rm NS}}\right)^{-1},
\\
F_{{\rm TD},{\rm tr}} &=
\frac{F_{{\rm TD},{\rm ob}}}{1-\mathcal{B}(R_{\rm TD})}
= \frac{cGM_{\rm NS}}{\kappa D^2}
\left(1-\frac{2GM_{\rm NS}}{c^2R_{\rm TD}}\right)^{1/2}.
\end{align*}

\subsubsection{Case study implementation}
We employ the same blocking function $\mathcal{B}(R)$ utilized in our previous photosphere evolution constraints analysis.
This particular blocking function was selected as it demonstrates consistency with our earlier results in Sect.~\ref{sec3.4.2:photosphereevolution}.

For this exemplary case, we impose the condition $R_{\rm {TD}} = 1.1\,R_{\rm {NS}}$, representing a scenario where touchdown occurs at a radius slightly larger
than the neutron star surface. This choice reflects the possibility suggested by \citet{Steiner2010} that the photosphere may not contract precisely to the
neutron star surface.

We solve the above modified system numerically, accounting for the radius-dependent blocking factors.
Our numerical solution yields $M_{\rm{NS}}\approx1.17\,{\rm{M}_{\sun}}$ and $R_{\rm{NS}}\approx13.4\,\rm{km}$, yielding values consistent with canonical neutron star parameters.
The corresponding touchdown radius is $R_{\mathrm{TD}} \approx 14.7\,\mathrm{km}$, with blocking factors $\mathcal{B}(R_{\mathrm{NS}}) \approx 0.818$
and $\mathcal{B}(R_{\mathrm{TD}}) \approx 0.792$. Applying the peak-to-touchdown flux ratio $f_{\rm{peak/TD}}=2.0$ to Eq.~(\ref{eq7:fpeaktd}),
we determine $R_{\rm{peak}} \approx 39.6\,\rm{km}$ with $\mathcal{B}(R_{\rm{peak}}) \approx 0.619$. This yields a photosphere contraction speed of approximately
$12\ {\rm km\ s^{-1}}$, which remains well within the subsonic regime.

\subsection{Equation of state implications: Comparison with theoretical models}
Table \ref{table4:eosmodels} presents the nine different EoS models used for comparison with our mass--radius constraints. These models represent
different theoretical approaches to describing dense matter in neutron star cores.

The first six models are based on normal nuclear matter composed of nucleons, with the H3 model additionally incorporating hyperons. In contrast, the three
SQM models (SQM1, SQM2, SQM3) include the effects of deconfined quarks and represent potential quark star configurations.

Our results demonstrate a clear transition in the viability of different EoS models depending on the blocking factor:

Without blocking correction $(\mathcal{B}=0)$: the anomalously small mass and radius estimates ($M\sim0.2-0.4\,\rm{M_{\sun}}$, $R\sim6-9\,\rm{km}$) can only
be accommodated by the quark matter models SQM1, SQM2, and SQM3. All normal nuclear matter models predict significantly larger neutron star parameters that are
inconsistent with the uncorrected observational constraints. 

With significant blocking $(\mathcal{B}\gtrsim0.8)$: the corrected mass and radius estimates become consistent with canonical neutron star values and are compatible with
standard nucleonic (and nucleonic+hyperon) equations of state. In this regime, the specific self-bound strange-quark-matter (SQM) models considered here are
no longer required to explain the observations. We note, however, that mass--radius constraints alone cannot exclude the presence of deconfined quarks in the core
(e.g., hybrid stars or phase transitions) if the resulting macroscopic $M$--$R$ relation overlaps that of nucleonic equations of state.

This result is particularly significant given the existence of neutron stars with masses exceeding 2\,$M_{\sun}$ \citep{Demorest2010,Antoniadis2013}.
All viable EoS models must be able to support such massive configurations, which disfavors models that are too soft at high density
(including some quark-matter parameterizations), but does not uniquely determine the microscopic composition of the core.

\section{Discussion}
Our reanalysis of the neutron star in 4U 1746-37 demonstrates that introducing a blocking factor $\mathcal{B}$ to quantify systematic reduction in observed flux
due to the high-inclination geometry can reconcile the anomalous mass and radius predicted by \citet{Li2015} with canonical neutron star properties.
Our blocking factor approach yields three key results:
\begin{description}
\item [Canonical mass--radius recovery:] significant blocking ($\mathcal{B}$ $\gtrsim$ 0.8) reconciles the anomalous estimates with standard neutron star parameters
(M $\sim$ 1.6--2.1 M$_{\odot}$, R $\sim$ 10--13 km).

\item [Flux ratio resolution:] the radius-dependent blocking naturally explains the observed peak-to-touchdown ratio ($\sim$ 2.0) that exceeds general
relativistic limits.

\item [Subsonic evolution:] photospheric contraction speeds (5--12 km s$^{-1}$) remain well below the sound speed, confirming physical consistency.
\end{description}
However, several important methodological considerations and assumptions underlying our blocking factor approach require careful examination.
While our results successfully reconcile the anomalous mass--radius estimates with canonical neutron star properties, the interpretation relies on specific
choices regarding burst analysis techniques, atmospheric physics parameters, and geometric modeling assumptions that may introduce systematic uncertainties.
Furthermore, our approach operates within the framework of the touchdown method for PRE burst analysis, and does not thoroughly consider alternative approaches
such as the cooling tail method that have gained prominence in recent neutron star studies. We address these issues in the following subsections.

\subsection{The cooling tail method}
\label{subsec:4.1}
\subsubsection{Historical context and methodological differences}
The estimation of neutron star masses and radii from PRE bursts has been approached through two primary methodologies with distinct approaches and assumptions.
The touchdown method, employed in this work and dating back to the foundational studies by \citet{Lewin1993}, identifies the moment when the expanded photosphere
contracts back to the neutron star surface, using touchdown flux as a proxy for the Eddington limit and the effective emitting area during the cooling phase.

The cooling tail method, developed more recently by \citet{Suleimanov2011ApJ} and refined in subsequent studies \citep{Poutanen2014,Nattila2016,Suleimanov2017},
focuses on the spectral evolution during the cooling phase after touchdown. This approach leverages sophisticated neutron star atmospheric models that predict
the evolution of the color-correction factor $f_{\rm c}$ as a function of luminosity, comparing these predictions with observed spectral changes to constrain
neutron star properties.

\subsubsection{Spectral state dependencies and selection criteria}
A fundamental distinction between these two methods lies in their applicability to different observational conditions.
The cooling tail method requires that bursts occur during the low, hard spectral state of the source, when persistent accretion flux is minimal and
atmospheric model predictions are expected to be most reliable. Under these conditions, the effective area $A$ is predicted to evolve with flux in a specific
manner that reflects the varying color-correction factor, with $f_{\rm c}$ evolving from approximately 2.0 when luminosity reaches the Eddington limit
to 1.3--1.4 during the cooling tail phase.

However, most bursts analyzed using the touchdown method, including those from 4U 1746-37, occur during the soft, high accretion rate state when the
persistent flux is dominated by the accretion disk. In this regime, the effective area shows little to no evolution with flux, which appears inconsistent
with atmospheric model predictions. This discrepancy has led proponents of the cooling tail method to argue that such bursts cannot provide reliable
mass and radius constraints, as the observed spectral evolution does not match theoretical expectations for neutron star atmospheres \citep{Kajava2014}.

In the same soft-state regime, however, the observed touchdown flux need not be identified directly with the intrinsic Eddington flux. In a high-inclination system
such as 4U~1746$-$37, disk obscuration and reprocessing can also modify the flux measured at touchdown, just as they complicate the cooling-tail analysis.
Our use of the touchdown method in this work should therefore not be understood as assuming that the observed touchdown flux is unaffected by the accretion flow.
Rather, the working hypothesis is that the observed touchdown flux is a geometrically attenuated proxy for the intrinsic flux, and that the dominant first-order effect of
the accretion geometry can be represented through a net blocking factor, $\mathcal{B}$.

\subsubsection{Implications for 4U 1746-37}
For 4U 1746-37 specifically, \citet{Li2015} found that while the source's $A^{-1/4}$ flux correlation matches atmospheric model predictions at high flux,
it deviates significantly at low flux where the cooling tail method would typically be applied. Additionally, the relatively low blackbody temperatures
$(< 2.5 \rm{keV})$ observed in 4U 1746-37's cooling tails may fall outside the regime where strong color-correction factor evolution is expected
\citep{Guver2012a,Guver2012b}.

The high accretion rate environment characteristic of 4U 1746-37's soft state may fundamentally alter atmospheric structure in ways that affect both the
intrinsic color-correction factor evolution and the observed spectral properties. Furthermore, as demonstrated by \citet{Suleimanov2018}, accretion heating
can significantly reduce blackbody normalization compared to nonheated atmospheres, potentially introducing systematic biases in mass--radius estimation if not accounted for.

Our blocking factor approach provides a first-order framework for interpreting these apparent inconsistencies between methods.
In this picture, the soft-state geometry does not preserve the standard touchdown assumption in a literal sense; instead, the observed touchdown flux is treated as a reduced
proxy for the intrinsic Eddington-limited flux, $F_{\rm TD,ob} = (1-\mathcal{B})\,F_{\rm TD,tr}$, where $\mathcal{B}$ parameterizes the net loss of directly visible photospheric emission
due to geometric obscuration and related flux-reducing effects. The purpose of this phenomenological correction is not to provide a complete physical description of the burst environment,
but rather to test whether a net multiplicative attenuation is sufficient to reconcile the inferred mass--radius constraints with canonical neutron star values.
In this sense, the same soft-state geometry that limits the applicability of the cooling-tail method also motivates the blocking-factor treatment adopted here.

At the same time, this simplified treatment has clear limitations. A single multiplicative factor cannot capture energy-dependent reprocessing, spectral-shape distortions,
azimuthal asymmetries, or time-dependent three-dimensional disk structure during the burst. A fully self-consistent treatment would require coupled neutron star atmosphere
and accretion-disk radiative-transfer calculations. We therefore regard the blocking-factor formalism as an exploratory geometric correction, useful for assessing the plausibility
of soft-state flux attenuation, rather than as a complete physical model.

\subsection{Color-correction factor and atmospheric physics}
\label{subsec:4.2}
\label{sec:colorcorrection}
Theoretical neutron star atmospheric models predict that the color-correction factor $f_{\rm c}$ should evolve significantly during PRE bursts, potentially
ranging from $f_{\rm c} \approx 2.0$ at the Eddington limit to $f_{\rm c} \approx 1.3-1.4$ during the cooling tail phase. This evolution reflects changes in
the atmospheric structure and opacity as the luminosity decreases from the Eddington limit following the predictions of detailed atmospheric modeling
\citep{Suleimanov2011ApJ,Suleimanov2017}.

In our analysis, we adopted a constant value $f_{\rm c} = 1.35 \pm 0.05$.
While this assumption simplifies the calculations and aligns with the touchdown method's assumptions, it contrasts with the evolving $f_{\rm c}$ predicted by
atmospheric models. Nevertheless, there are further considerations regarding the $f_{\rm c}$ evolution in this context, as detailed below under
observational constraints, accretion effects, and geometric considerations.

\begin{description}
  \item [Observational constraints:] for 4U 1746-37, the limited temperature range $(< 2.5 \rm{keV})$ during the cooling phase suggests that $f_{\rm c}$
  evolution may be less pronounced than in higher temperature bursts. As demonstrated by \citet{Guver2012a,Guver2012b}, the color-correction factor shows
  minimal dependence on temperature when blackbody temperatures remain below 2.5 keV, which is precisely the regime observed in 4U 1746-37's cooling tails.
  \item [Accretion effects:] the high accretion rate environment in 4U 1746-37, evidenced by its predominantly soft spectral state, may alter the atmospheric
  structure in ways that could potentially affect the $f_{\rm c}$ evolution. \citet{Suleimanov2018} demonstrated that accretion-heated neutron star atmospheres
  exhibit substantially different spectral properties compared to nonheated atmospheres, potentially modifying the expected $f_{\rm c}$ evolution patterns.
  \item [Geometric considerations:] the apparent tension between fixed $f_{\rm c}$ assumptions and evolving atmospheric models may be partially resolved by
  considering the geometric effects central to our analysis. If significant portions of the neutron star surface are blocked throughout the burst evolution,
  the observed spectral properties may not accurately reflect the intrinsic atmospheric conditions, making the traditional $f_{\rm c}$ evolution predictions
  less applicable to high-inclination systems like 4U 1746-37.
\end{description}

This interpretation could provide a potential bridge between the minimal color-correction evolution in 4U 1746-37 and the theoretical expectations of significant
atmospheric evolution during PRE bursts. Future work incorporating both detailed atmospheric modeling and three-dimensional geometric effects will be essential
to fully resolve these apparent inconsistencies.

\subsection{Limitations and systematic uncertainties}
\label{sec:4.3}
\subsubsection{Geometric model dependencies}
Our blocking factor calculations rely on simplified accretion disk models \citep{Shakura1973,Narayan1994} that, while physically motivated, should be considered
only illustrative. More sophisticated models incorporating radiative transfer and three-dimensional structure would provide better physical foundations for
blocking factor calculations. 

A further check on the plausibility of large blocking factors comes from the known geometry of 4U~1746$-$37. The source is a dipping LMXB, which independently indicates
a high-inclination viewing geometry. In the standard geometric picture for LMXBs, dips without eclipses are generally expected for inclinations
of order $\sim 60^\circ$--~$75^\circ$, whereas eclipses occur only at somewhat higher inclinations \citep{Frank1987}. Since 4U~1746$-$37 is observed as a dip source
rather than as a persistently eclipsing system \citep{Balman2009}, our high $\mathcal{B}$ solutions may be regarded as plausible for the upper end of the allowed high-inclination,
non-eclipsing regime, provided that the inner accretion flow is sufficiently vertically extended or azimuthally structured.

In addition, the simple multiplicative correction adopted in this work should be regarded as phenomenological. It captures the net attenuation of the directly visible burst emission,
but does not explicitly model energy-dependent reprocessing, angular redistribution of radiation, or spectral distortions introduced by the accretion flow. Although we parameterize
the blocking effect using a single factor $\mathcal{B}$, this quantity should not be interpreted too literally as a purely geometric scalar. In a disk-dominated soft-state system,
energy-dependent scattering, spectral distortion, anisotropic redistribution, and partial reprocessing may affect the observed touchdown flux and apparent area in different ways.
The resulting $M$--$R$ shifts therefore need not reduce to a simple rigid displacement under a single multiplicative correction; instead, the allowed regions may broaden, deform,
or shift quantitatively. We thus regard the present $\mathcal{B}$-based treatment as a first-order phenomenological diagnostic, with the highest values ($\mathcal{B}\sim 0.9$)
representing upper-end illustrative cases rather than uniquely established properties of 4U~1746$-$37.

\subsubsection{Parameter transformation biases}
The transformation from observable quantities $(F_{\rm {TD}}, A)$ to physical parameters $(M, R)$ can introduce systematic biases, particularly when the
Jacobian of the transformation approaches zero near the general relativistic limit. Our Monte Carlo approach partially addresses this concern by operating
with probability distributions rather than point estimates. Future work could explore more sophisticated Bayesian approaches that operate directly in
$(M, R)$ space.

\section{Conclusion}
The blocking factor correction is intentionally minimalistic, incorporating observed flux-reducing effects such as the disk geometry, anisotropic emission
patterns, and reflection effects into a single parameter. This parameter rescales the observed flux and effective area without requiring
modifications to spectral shape assumptions. However, several methodological choices in our analysis warrant acknowledgment.
We adopted simplified disk geometries, employed a constant color-correction factor, and utilized the touchdown method's standard framework for identifying key burst phases.
Each of these assumptions may introduce systematic uncertainties that merit investigation through more sophisticated modeling
approaches. Additionally, parameter transformations near relativistic compactness limits can potentially bias posterior distributions, suggesting that
future work should consider Bayesian inference methods operating directly in mass--radius space while incorporating radius-dependent blocking functions
and detailed atmospheric models.

The natural progression of this work involves developing self-consistent forward models that couple time-dependent neutron star atmospheres with
three-dimensional radiative transfer through realistic accretion disk geometries. Similar models would predict blocking factors $\mathcal{B}(R_{\rm ph}, \theta_{\rm in})$,
spectral evolution, and burst light curves simultaneously. From an observational perspective, coordinated campaigns combining burst timing and spectroscopy
with independent geometric constraints, such as orbital inclination measurements from dipping behavior, disk thickness indicators, and spectral state
classifications, would provide stronger priors on blocking factors.

This work reframes 4U 1746-37 as a constructive example in neutron star mass--radius measurements. When geometric effects are explicitly incorporated through
blocking factor corrections, the apparent inconsistencies among photospheric radius expansion burst observations, flux ratio anomalies, and photospheric
contraction kinematics are resolved within a framework consistent with canonical neutron star structure.

\begin{acknowledgements}
We thank the referee for the careful reading and constructive comments, which have significantly improved the manuscript.\\
MK acknowledges the support from the Institute for Basic Science (IBS) of the Republic of Korea (Grants No. IBS-R031-D1).
YMK was partially supported by the National Research Foundation of Korea (NRF) grant (No. RS-2022-NR072453) and the KASI research project (No. 2026183200) funded by the Korea government (MSIT).
C.-H.L. was supported by the National Research Foundation of Korea (NRF) grant funded by the Korea government (No. RS-2023-NR076639).
KK was supported by the National Research Foundation of Korea (NRF) grants (2022R1F1A1073890, RS-2025-00516133) and Institute of Information \& communications Technology Planning \& Evaluation (IITP) grant RS-2021-II212068 (Artificial Intelligence Innovation Hub) funded by the Korea government (MSIT).\\
KHS and MK contributed equally to this work as co-first authors by carrying out the analytic calculations and the Monte Carlo sampling, respectively.
\end{acknowledgements}

\bibliographystyle{aa}
\bibliography{rmr}

\begin{appendix}
\begin{sidewaystable*}
   \section{Disk structure models and parameters}
   \caption{Scale height, density, and temperature parameters for disk models}
   \label{tableA1:disk_structure}
   \centering
   \begin{tabularx}{0.8\textwidth}{cYYY}
      \toprule\toprule
      & $z_0~(\mathrm{cm})$ & $\rho_0~(\mathrm{g~cm^{-3}})$ & $T~(\mathrm{K})$ \\
      \midrule
      \\
      {\it Case}~1$^{a}$ &
      $7.8 \times 10^5 ~(\frac{\dot{M}}{0.25 \dot{M}_{\rm Edd}}) \times [1-(\frac{R}{R_0})^{-1/2}]$ &
      $9.9 \times 10^{-5} ~(\frac{\alpha_v}{0.1})^{-1} (\frac{\dot{M}}{0.25 \dot{M}_{\rm Edd}})^{-2} \newline \times (\frac{M}{1.4M_{\sun}})^{-1/2} (\frac{R}{R_0})^{3/2} [1-(\frac{R}{R_0})^{-1/2}]^{-2}$ & 
      $3.3 \times 10^7 ~(\frac{\alpha_v}{0.1})^{-1/4} \times (\frac{M}{1.4M_{\sun}})^{1/8} (\frac{R}{R_0})^{-3/8}$ \\
      \\
      {\it Case}~2$^{b}$ &
      $1.4 \times 10^4 ~ (\frac{\alpha_v}{0.1})^{-0.1} (\frac{\dot{M}}{0.25 \dot{M}_{\rm Edd}})^{1/5} \newline \times (\frac{M}{1.4M_{\sun}})^{-0.35} (\frac{R}{R_0})^{1.05} [1-(\frac{R}{R_0})^{-1/2}]^{1/5}$ &
      $16~ (\frac{\alpha_v}{0.1})^{-0.7} (\frac{\dot{M}}{0.25 \dot{M}_{\rm Edd}})^{0.4} \newline \times (\frac{M}{1.4M_{\sun}})^{0.55} (\frac{R}{R_0})^{-33/20} [1-(\frac{R}{R_0})^{-1/2}]^{0.4}$ &
      $2.4 \times 10^8 ~(\frac{\alpha_v}{0.1})^{-0.2} (\frac{\dot{M}}{0.25 \dot{M}_{\rm Edd}})^{0.4} \newline \times (\frac{M}{1.4M_{\sun}})^{0.3} (\frac{R}{R_0})^{-0.9} [1-(\frac{R}{R_0})^{-1/2}]^{0.4}$ \\
      \\
      {\it Case}~3$^{c}$ &
      $8.2 \times 10^3 ~(\frac{\alpha_v}{0.1})^{-0.1} (\frac{\dot{M}}{0.25 \dot{M}_{\rm Edd}})^{0.15} \newline \times (\frac{M}{1.4M_{\sun}})^{-3/8} (\frac{R}{R_0})^{9/8} [1-(\frac{R}{R_0})^{-1/2}]^{0.15}$ &
      $84 ~ (\frac{\alpha_v}{0.1})^{-0.7} (\frac{\dot{M}}{0.25 \dot{M}_{\rm Edd}})^{0.55} \newline \times (\frac{M}{1.4M_{\sun}})^{5/8} (\frac{R}{R_0})^{-15/8} [1-(\frac{R}{R_0})^{-1/2}]^{0.55}$ &
      $7.9 \times 10^7 ~ (\frac{\alpha_v}{0.1})^{-0.2} (\frac{\dot{M}}{0.25 \dot{M}_{\rm Edd}})^{0.3} \newline \times (\frac{M}{1.4M_{\sun}})^{1/4} (\frac{R}{R_0})^{-3/4} [1-(\frac{R}{R_0})^{-1/2}]^{0.3}$ \\
      \\
      {\it Case}~4$^{d}$ &
      $6.7 \times 10^{5} ~ (\frac{R}{R_0})$ &
      $1.1 \times 10^{-4} ~ (\frac{\alpha_v}{0.1})^{-0.1} (\frac{\dot{M}}{0.25 \dot{M}_{\rm Edd}}) \newline \times (\frac{M}{1.4M_{\sun}})^{-1/2} (\frac{R}{R_0})^{-3/2}$ &
      $3.2 \times 10^{11}~ (\frac{M}{1.4M_{\sun}}) (\frac{R}{R_0})^{-1}$ \\
      \\
      \bottomrule
      \\
   \end{tabularx}
   \vfill
   \raggedright
   $^{a}$~Thin-disk model from \citet{Shakura1973}, Region  a): radiation pressure dominated, Thomson scattering opacity $P_{\rm r} \gg P_{\rm g}$ and $\sigma_{\rm T} \gg \sigma_{\rm ff}$.\\
   $^{b}$~Thick-disk model from \citet{Shakura1973}, Region  b): gas pressure dominated, Thomson scattering opacity $P_{\rm r} \ll P_{\rm g}$ and $\sigma_{\rm T} \gg \sigma_{\rm ff}$.\\
   $^{c}$~Thick-disk model from \citet{Shakura1973}, Region  c): gas pressure dominated, free-free opacity $P_{\rm r} \ll P_{\rm g}$ and $\sigma_{\rm T} \ll \sigma_{\rm ff}$.\\
   $^{d}$~Advection-dominated accretion flow (ADAF) from \citep{Narayan1994} with no radiative cooling ($f=1$) and $\gamma=4/3$.\\
   
   \tablefoot{Parameters are expressed in terms of accretion rate $({\dot{M}})$, central mass ($M$), radius ($R$), and viscosity $(\alpha_v)$. Scale height $z_0$
   represents half-thickness, $\rho_0$ is volume mass density,\\ and T is temperature. For thick-disk models (Cases 2-4), temperature is derived from sound speed;
   for the thin-disk model (Case 1), it is derived from energy density. The Eddington\\ accretion rate $\dot{M}_{\rm Edd}$ and reference radius $R_0$ are
   treated as constants independent of stellar mass $M$ for consistency across models.}
  \tablefoot{Usage in main analysis: in the geometric analysis presented in Sect.~2, the thin-disk model corresponds to the standard
  Shakura-Sunyaev disk (Case 1), \\while the thick-disk model is based on the advection-dominated accretion flow (ADAF) solution (Case 4).}
\end{sidewaystable*}
\clearpage

\begin{figure*}[!ht]
   \centering
   \begin{subfigure}{0.46\textwidth}
      \centering
      \includegraphics[width=\linewidth]{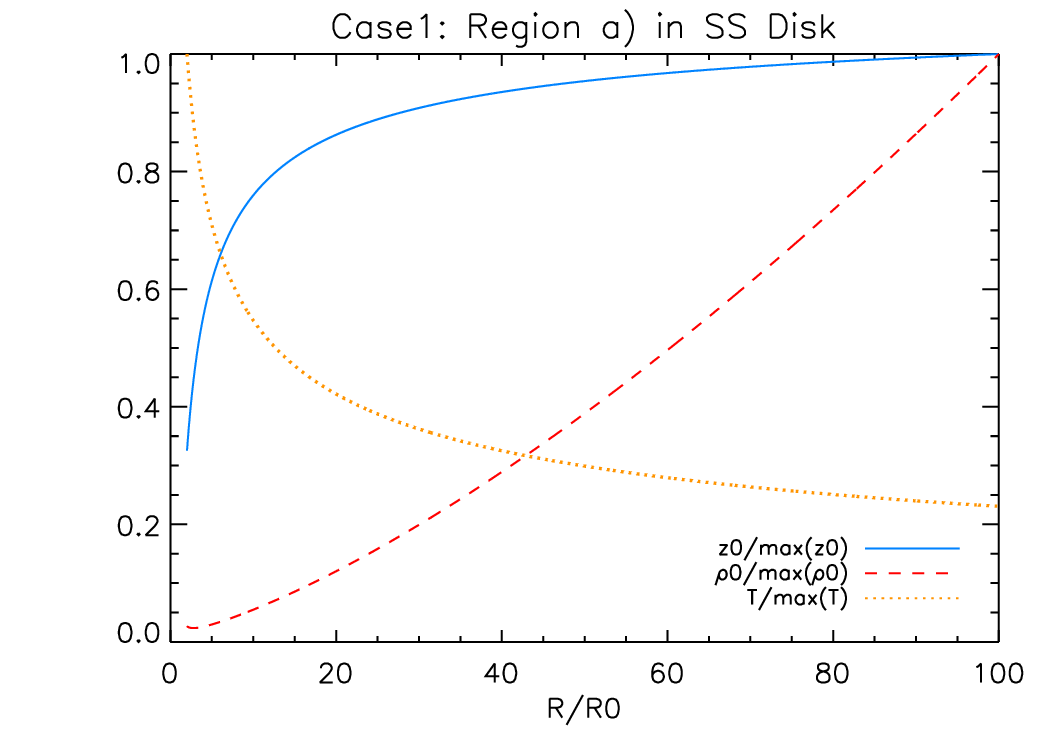}
      \label{thin_a}
   \end{subfigure}
   \hspace{0.06\textwidth}
   \begin{subfigure}{0.46\textwidth}
      \centering
      \includegraphics[width=\linewidth]{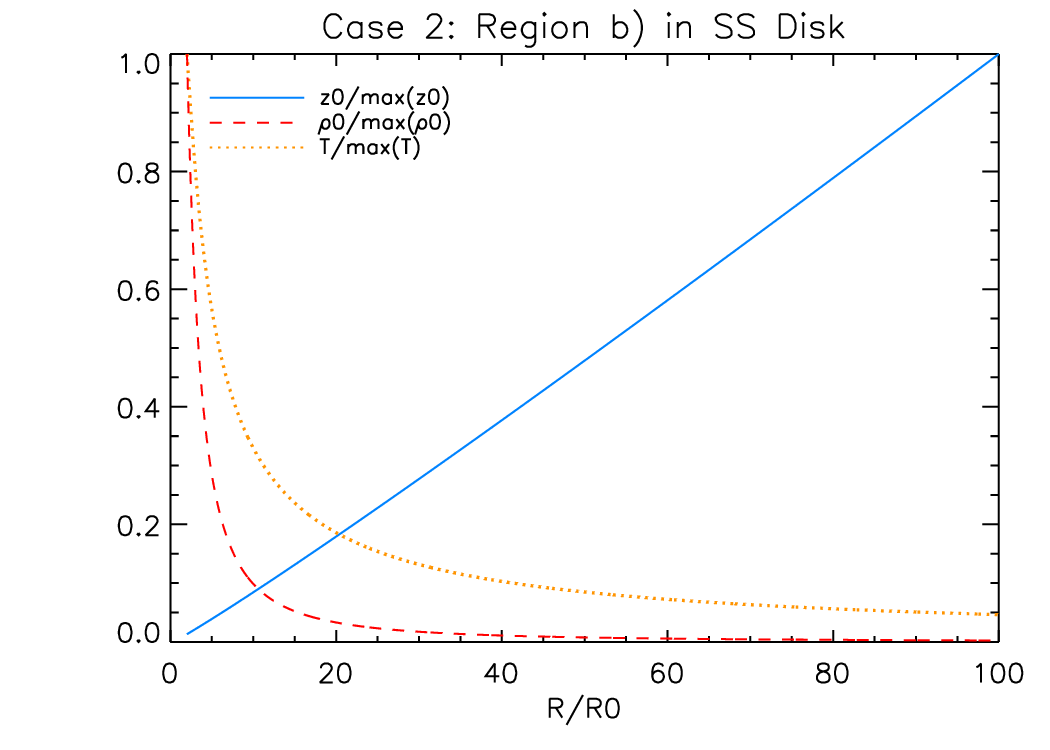}
      \label{thin_b}
   \end{subfigure}
   \begin{subfigure}{0.46\textwidth}
      \centering
      \includegraphics[width=\linewidth]{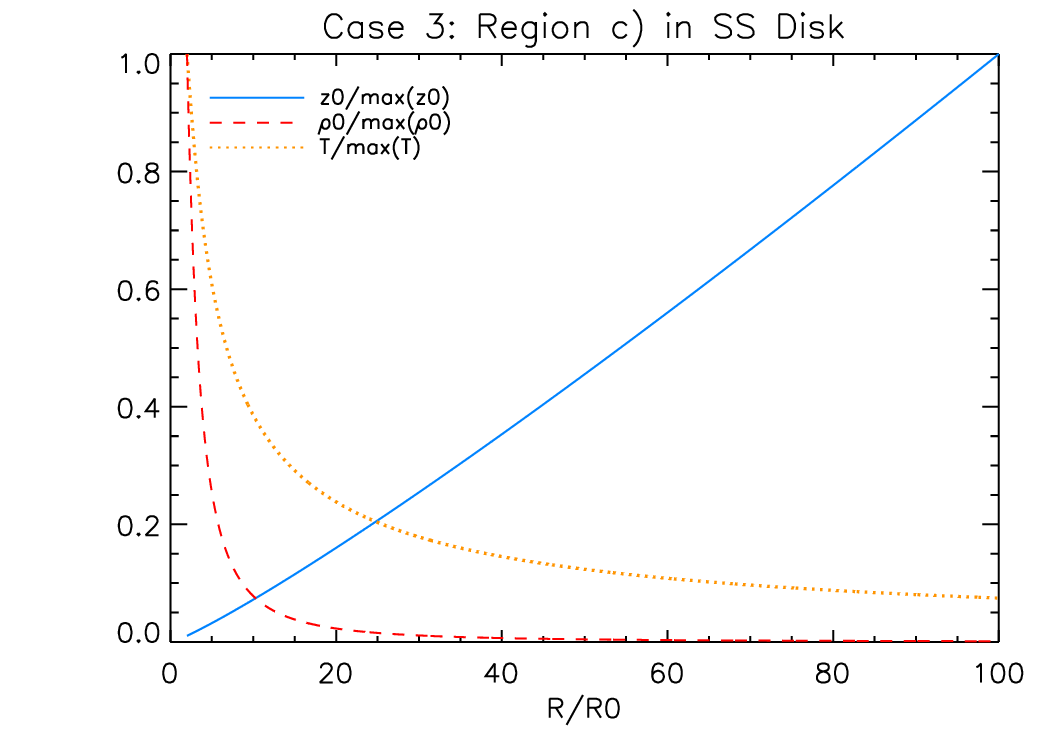}
      \label{thin_c}
   \end{subfigure}
   \hspace{0.06\textwidth}
   \begin{subfigure}{0.46\textwidth}
      \centering
      \includegraphics[width=\linewidth]{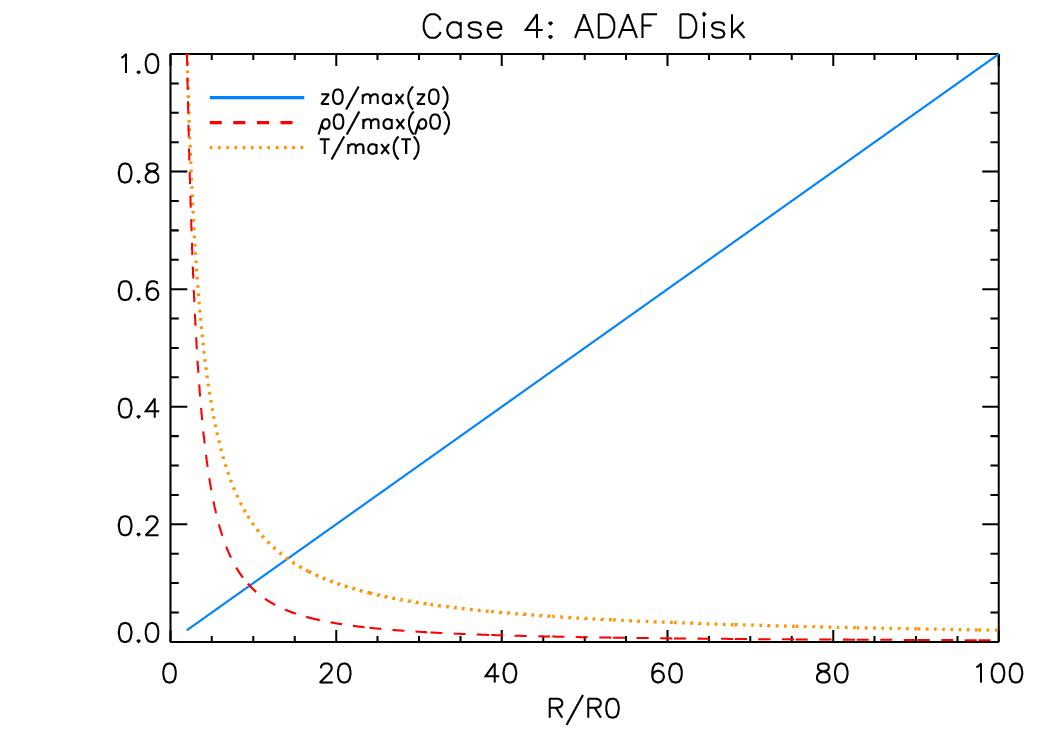}
      \label{thick_d}
   \end{subfigure}
   \caption{Radial profiles of physical parameters for the four accretion disk models used in blocking factor calculations. Top left: Case 1 (thin disk,
   \citet{Shakura1973} Region a)). Top right: Case 2 (thick disk, \citet{Shakura1973} Region b)). Bottom left: Case 3 (thick disk, \citet{Shakura1973} Region c)).
   Bottom right: Case 4 (ADAF thick disk, \citet{Narayan1994}). Each panel shows the normalized radial dependence of scale height $z_0$ (solid line),
   temperature $T$ (dashed line), and volume density $\rho_0$ (dotted line) as functions of the normalized radial coordinate $\rm{R}/\rm{R_0}$, where $\rm{R_0}\approx12$ km.
   All quantities are scaled to their respective maximum values within this radial range to facilitate comparison between models. The disk models span a wide range of physical
   conditions, from geometrically thin configurations (Case 1) to highly inflated thick-disk structures (Case 4). Standard parameters $(\alpha_v)=0.1$ and
   ${\dot{M}}$=$0.25\dot{M}_{\rm Edd}\approx9.1\times10^{-9}\,\rm{M_{\sun}}\,\rm{yr}^{-1}$ are adopted throughout. See Table \ref{tableA1:disk_structure} for
   complete parameter expressions and model specifications.}
   \label{fig_disk_cases}
\end{figure*}

\end{appendix}

\end{document}